%% file: manuscript-preprint.tex
\documentclass[11pt]{article}

\usepackage{arxiv} 

\usepackage{algorithm,algpseudocode}
\usepackage{amsbsy,amsthm,amsmath,amssymb}
\usepackage{appendix}
\usepackage{booktabs}
\usepackage{graphicx}
\usepackage{natbib}
\usepackage{xcolor}

\title{Feature Selection for Vertex Discriminant Analysis}
\author{Alfonso Landeros$^{2}$, Tong Tong Wu$^{1}$, Kenneth Lange$^{2,3,4}$
    \\
    \\
    Department of Biostatistics and Computational Biology$^{1}$ \\
    University of Rochester \\
    Rochester, New York 14642-0630 \\
    \\
    Departments of Computational Medicine$^{2}$, \\
    Human Genetics$^{3}$, and Statistics$^{4}$ \\
    University of California \\
    Los Angeles, CA 90095\\
    Phone: 310-206-8076 \\
    E-mail klange@ucla.edu \\
    \\
    \\
    \\
    Research supported in part by
    \\ USPHS grants GM53275 and HG006139 
    \\ and NSF grant CCF-1934962.
}

\input{preamble.tex}

\begin{document}
\maketitle

\begin{abstract}
    We revisit vertex discriminant analysis (VDA) from the perspective of proximal distance algorithms.
    By specifying sparsity sets as constraints that directly control the number of active features, VDA is able to fit multiclass classifiers with no more than $k$ active features.
    We combine our sparse VDA approach with repeated cross validation to fit classifiers across the full range of model sizes on a given dataset.
    Our numerical examples demonstrate that grappling with sparsity directly is an attractive approach to model building in high-dimensional settings.
    Applications to kernel-based VDA are also considered.
\end{abstract}
\label{sec:intro}
Support vector machines (SVM) stand out in the universe of classification algorithms for their simplicity, interpretability, speed, and theoretical guarantees \citep{smola1998learning,vapnik1999nature}.
The circle of SVM algorithms have been bolstered by fundamental advances in statistical theory and computer hardware but also challenged by the growing size of modern datasets.
The Achilles heal of SVMs is their limitation to binary classification.
Strategies to overcome this limitation, most notably one-versus-rest and one-versus-one, rely on multiple SVMs to discriminate between classes.
This necessarily increases time to fit a model, obscures statistical interpretation, and violates the principle of parsimony.
Vertex discriminant analysis (VDA) \citep{lange2008mm} addresses this particular weakness of SVMs.
Our subsequent paper \citep{wu2010multicategory} derives a fast cyclic coordinate descent algorithm and attacks feature selection by a combination of Euclidean group and lasso penalties.
Our second followup paper \citep{wu2012nonlinear} generalizes VDA to nonlinear discriminant analysis by the kernel trick.

The current paper revisits the problem of feature selection in VDA from the perspective of distance to set penalties.
This allows us to incorporate a spectrum of constraints, with particular emphasis on sparsity sets.
The experiments recorded here demonstrate that the direct control of sparsity is highly effective in building parsimonious models on both simulated and real-world datasets. 

\section{Proposed Method}
\label{sec:method}
As a prelude to introducing a generic method for feature constraints, let us review VDA with $\epsilon$-insensitive pseudo-distances and derive a quadratic surrogate that both simplifies the associated empirical loss and converts minimization to iterative least squares.
Next, we present a distance penalized version of VDA that reduces constrained minimization to unconstrained minimization.
In particular, we focus on distance to sparsity sets as a mechanism for directly enforcing parsimony.
We then present two competitive MM algorithms designed with high-dimensional optimization in mind.
Finally, we comment on practical issues, including initialization strategies, convergence criteria, and hyperparameter tuning.

Before delving into specifics, let us clarify our notation. We restrict our attention to problems formulated over Euclidean space $\Real^{n}$.
Vectors are written in lowercase boldface, for instance $\by$, and matrices are written in uppercase boldface, for instance $\bX$.
The notation $\|\cdot\|$ indicates the Euclidean norm on the ambient space.
In some cases we decorate a norm with a subscript, for example the Frobenius norm $\|\cdot\|_{F}$, to avoid ambiguity.
The superscript $\top$ indicates a vector or matrix transpose.
The lowercase letter $k$ denotes the number of selected parameters; equivalently, the lowercase letter $s = 1-k/p$ is reserved for the sparsity level.
The maximum number of parameters $p$ is understood from context.
Finally, the letters $m$ and $t$ are reserved for inner iterations and outer iterations, respectively. 
For example, $\bB_{m}$ is an estimate at iteration $m$ to a problem $\min_{\bB}~f_{\rho(t)}(\bB)$ parameterized by a penalty coefficient $\rho(t)$, whereas $\bB_{t}$ denotes one of its solutions, either exact or approximate, at level $t$.

\subsection[Majorizing Epsilon-Insensitive Empirical Risk]{Majorizing $\epsilon$-Insensitive Empirical Risk}

Given a dataset with $n$ samples, $p$ predictors, and $c$ categories, the goal in VDA is to minimize the empirical risk
\begin{eqnarray*}
f(\bB) & = &  \frac{1}{n} \sum_{i=1}^{n} R(\by_{i}, \bx_{i}, \bB).
\end{eqnarray*}
Here $\bx_{i} \in \mathbb{R}^{p}$ is the feature vector for sample $i$,  $\by_{i} \in \mathbb{R}^{c-1}$ is the sample's label encoded as a vertex of a regular simplex in $\mathbb{R}^{c-1}$, and the entries of $\bB=(b_{jk})$ are regression coefficients to be estimated.
When $c=2$, $3$, or $4$, the simplex is $[-1,1]$, an equilateral triangle, or a regular tetrahedron centered around the origin, respectively.
The predicted value for sample $i$'s class vertex $\by_i$ is $\bB^\top\bx_i$.
We collect the rows $\bx_i^\top$ into a feature  matrix $\bX$ and the rows $\by_i^\top$ into a class matrix $\bY$.
To fit $\bX\bB$ to $\bY$, we focus attention on the risk function 
\begin{eqnarray}
    \label{eq:basic-model}
    f(\bB)
     & = & \frac{1}{2n} \sum_{i=1}^{n} 
     \max\{0, \|\by_{i} - \bB^\top\bx_i\|_{2} - \epsilon
    \}^{2},
\end{eqnarray}
which is the square of $\epsilon$-insensitive pseudo-distance. The residual $\by_i-\bB^\top\bx_i$ for sample $i$ incurs no loss when $\bB^\top\bx_i$ lies within a Euclidean $\epsilon$-ball around its assigned vertex $\by_{i}$. Otherwise, the residual contributes the reduced  loss $(\|\by_{i} - \bB^\top\bx_i\| - \epsilon)^{2}$ to the risk. If $\mathcal{Y}$ denotes the collection of $c$ vertices and $\hat{\bB}$ the estimated coefficient matrix, then an unassigned feature vector $\bx$ is assigned to a class through the map 
\begin{eqnarray*}
\bx & \mapsto & \argmin_{\by \in \mathcal{Y}} \|\by - \hat{\bB}^\top\bx\|,
\end{eqnarray*}
singling out the closest vertex. Inclusion of an intercept can be beneficial when training class numbers are unbalanced, and easily is achieved by adding an intercept row to $\bB$ and a corresponding constant column $\bone$ to $\bX$. 

Squared $\epsilon$-insensitive pseudo-distances admit a quadratic surrogate that greatly simplifies minimization of the loss $f(\bB)$.
To identify the surrogate, consider the majorization
\begin{eqnarray*}
    \inf_{\bv} \{f_{1}(\bv) + f_{2}(\bu - \bv)\} & \le & f_{1}(\bv_{m}) + f_{2}(\bu - \bv_{m})
\end{eqnarray*}
of the infimal convolution of two function $f_1(\cdot)$ and $f_2(\cdot)$.
In addition to the indicated inequality for all $\bu$, majorization requires the tangency condition of equality when $\bu$ equals the anchor point $\bv_m$.
This is achieved by choosing $\bv_m$ to be the minimum point of the infimal convolution corresponding to $\bu_m$.
For the closed set $C=[-\epsilon,\epsilon]$, the $0/\infty$ indicator $f_1(\bv)=\delta_{C}(\|\bv\|)$ and the squared Euclidean distance $f_2(\bv) = \|\bv\|^2$ we deduce the majorization
\begin{eqnarray*}
    \max\{0, \|\bu\| - \epsilon \}^2
    &=&
    \min_{\bv} \delta_{[-\epsilon,\epsilon]}(\|\bv\|)
    +
    \|\bu - \bv\|^{2}
    \;\: \le \;\: \|\bu - \bv_{m}\|^{2}; \\
    \bv_{m} &=& \begin{cases}
        \bu_{m}, & \|\bu_{m}\| \le \epsilon \\
        \frac{\epsilon}{\|\bu_{m}\|} \bu_{m}, & \|\bu_{m}\| > \epsilon
    \end{cases}
\end{eqnarray*}
relevant to VDA. 

Applying this majorization to model \eqref{eq:basic-model} term-by-term, we arrive at the overall majorization
\begin{eqnarray*}
    f(\bB) &\le& \frac{1}{2n} \sum_{i=1}^{n} \|\br_{i} - \br_{mi}\|^{2}, \\
    \br_{i} &=& \by_{i} - \bB^\top\bx_i, \\
    \br_{mi} &=& \begin{cases}
        \by_{i} - \bB_m^\top\bx_i, & \|\by_{i}- \bB_{m}^\top\bx_i \| \le \epsilon \\
        \frac{\epsilon}{\|\by_{i} - \bB_{m}^\top\bx_i \|}(\by_{i} -\bB_{m}^\top\bx_i), & \|\by_{i} - \bB_{m}^\top\bx_i\| > \epsilon.
    \end{cases}
\end{eqnarray*}
Here the estimate at iteration $m$, denoted $\bB_{m}$, serves as an anchor for the surrogate function.
It is straightforward to verify that tangency occurs when $\bB = \bB_{m}$.
To make further progress, consider two cases concerning the squared term.
When the residual is modest, we have
\begin{eqnarray*}
    \br_{i} - \br_{mi}
    & = & (\by_{i} - \bB^{\top} \bx_{i}) - (\by_{i} - \bB_{m}^{\top} \bx_{i})
    \;\: =\;\: \bB_{m}^{\top} \bx_{i} - \bB^{\top} \bx_{i}.
\end{eqnarray*}
In the opposite case, the penalty argument reads
\begin{eqnarray*}
    \br_{i} - \br_{mi}
   & = & \left(1 - \frac{\epsilon}{\|\by_{i}^\top - \bx_i^\top\bB_{m}\|} \right) \by_{i}
    + \frac{\epsilon}{\|\by_{i}^\top - \bx_i^\top\bB_{m}\|} \bB_{m}^{\top} \bx_{i} - \bB^{\top} \bx_{i}.
\end{eqnarray*}
In either case, we have successfully isolated the linear predictor $\bB^\top \bx_i$.
This leads to the quadratic surrogate
\begin{eqnarray}
    \label{eq:basic-surrogate}
    g(\bB \mid \bB_{m})
    &=& \frac{1}{2n} \sum_{i=1}^{n} \|\bz_{mi} - \bB^{\top} \bx_{i}\|^{2}
    \;\:=\;\: \frac{1}{2n} \|\bZ_{m} - \bX \bB\|_{F}^{2},
\end{eqnarray}
where $\bX\bB$ predicts $\bY$ and $\bY$ is replaced by $\bZ_m$ with rows
\begin{eqnarray*}
    \bz_{mi}^\top
    &=& \begin{cases}
        \bx_i^\top\bB_{m} , & \|\by_{i} - \bB_m^\top \bx_{i}\| \le \epsilon \\
        w_{mi} \by_{i}^\top + (1-w_{m,i}) \bx_i^\top \bB_m, & \|\by_{i} - \bB_{m}^{\top} \bx_{i}\| > \epsilon
    \end{cases} \\
    w_{mi}
    &=& \frac{\|\by_{i} - \bB_{m}^{\top} \bx_{i}\|-\epsilon}{\|\by_{i} - \bB_{m}^{\top} \bx_{i}\|}.
\end{eqnarray*}

In summary, our surrogate can be interpreted as an ordinary sum of squares with shifted responses. Observe that the tangency requirement $g(\bB_{m} \mid \bB_{m}) = f(\bB_{m})$ implies
\begin{eqnarray*}
  f(\bB_m) 
   & = &  \frac{1}{2n} \sum_{i=1}^{n} \|\bz_{mi} - \bB_{m}^{\top} \bx_{i}\|^{2}\\
    & = & \frac{1}{2n} \sum_{ \|\by_{i} - \bB_{m}^{\top} \bx_{i}\| > \epsilon} \!\! w_{mi}^2 \|\by_{i} - \bB_{m}^{\top} \bx_{i}\|^{2}.
\end{eqnarray*}
Hence at iteration $m$, only samples with residual errors beyond $\epsilon$ contribute to the loss.
These samples tend to drive the subsequent update $\bB_{m+1}$.

\subsection{Feature Selection via Sparsity Sets}
\label{sec:feature}

Having derived a quadratic surrogate for the empirical risk \eqref{eq:basic-model}, we now focus on feature selection.
Following our previous work on proximal distance algorithms \citep{keys2019proximala}, we penalize $f(\bB)$ by the squared Euclidean distance of $\bB$ from the constraint set $S$.
This creates the objective
\begin{eqnarray}
    \label{eq:penalized}
    f_{\rho}(\bB) & = & f(\bB) + \frac{\rho}{2} \dist(\bB, S)^{2}
\end{eqnarray}
for $\rho>0$ large.
In binary classification $c=2$, so a natural choice for $S$ is
\begin{eqnarray*}
    S_{k} & = &  \Big\{\bb \in \Real^{p} : \sum_{j=1}^{p} 1_{\{b_{j} \neq 0\}} \le k\Big\}
\end{eqnarray*}
for $\bb=(b_j)$.
In words, the sparsity set $S_k$ consists of those vectors with at most $k$ nonzero components.
When an intercept $\bb_0$ is included in the model, the intercept is usually omitted in the sparsity set. 
Minimizing the penalized objective $f_{\rho}(\bB)$ with $S = S_{k}$ delivers a solution minimizing empirical risk with at most $k < p$ distinguishing features. 

When the number of classes $c > 2$, the most plausible way to define sparsity is through the number of active features.
A feature corresponds to a row $\bb_j$ of $\bB$, so we encode sparsity through the constraint set
\begin{eqnarray*}
    S_{k}^{\mathrm{row}} & = & \Bigg\{\bB= \begin{pmatrix} \bb_1 \\
    \cdots \\
    \bb_{p} \end{pmatrix} \in \Real^{p \times (c-1)} : \sum_{j=1}^{p} 1_{\{\|\bb_j\| = 0\}} \le k \Bigg\}.
\end{eqnarray*}
Thus, at most $k$ active rows (features) are allowed in the model.
Note that $S_{k}^{\mathrm{row}} \equiv S_{k}$ when $c = 2$.
Regardless of the constraint set $S$, distance majorization replaces $\dist(\bB,S)^2$ with $\|\bB - P_{S}(\bB_{m})\|_{F}^{2}$, where $P_{S}(\bB)$ denotes the Euclidean projection of $\bB$ onto $S$.
For a sparsity set, projection acts by ranking features according to their norms $\|\bb_j\|$ and then identifying the top $k$ features via a partial sort.
Omitted rows are then set to $\bzero^\top$.
The intercept row is ignored in projection.
In summary, projection onto a sparsity set amounts to computing order statistics on estimates of model coefficients and thresholding to select the highest ranking features.

In the special case $S=S_{k}^{\mathrm{row}}$, combining majorization \eqref{eq:basic-surrogate} with distance majorization leads to the overall quadratic surrogate
\begin{eqnarray}
    \label{eq:main-surrogate}
    g_{\rho}(\bB \mid \bB_{m})
    & = &
    \frac{1}{2n} \|\bZ_{m} - \bX \bB\|_{F}^{2} + \frac{\rho}{2} \|\bP_{m}-\bB\|_{F}^{2}
\end{eqnarray}
with the shifted response matrix $\bZ_m$ and projection $P_{S}(\bB_{m})$.
This convex quadratic serves as the main ingredient in deriving our novel proximal distance algorithms for VDA.

\subsection{Algorithms}

Let us briefly recall the MM principle \citep{lange2016mma} applied to the penalized objective (\ref{eq:penalized}).
Given an anchor point $\bB_{m}$, the surrogate (\ref{eq:main-surrogate}) majorizes (\ref{eq:penalized}) so that $f_{\rho}(\bB) \le g_{\rho}(\bB \mid \bB_{m})$ holds for every $\bB$ in the essential domain of the penalized objective.
At each iteration we select a potentially nonunique stationary point $\bB_{m+1} \in \argmin_{\bB}\, g_{\rho}(\bB \mid \bB_{m})$ minimizing the surrogate (\ref{eq:main-surrogate}) around the anchor point $\bB_m$.
These observations combined with the tangency condition reveal the chain of inequalities
\begin{eqnarray*}
    f_{\rho}(\bB_{m+1})
    \overset{\text{majorization}}{\le} g_{\rho}(\bB_{m+1} \mid \bB_{m})
    \overset{\text{definition}}{\le} g_{\rho}(\bB_{m} \mid \bB_{m})
    \overset{\text{tangency}}{=} f_{\rho}(\bB_{m}).
\end{eqnarray*}
Thus, minimizing surrogate (\ref{eq:main-surrogate}) guarantees the descent property in minimizing the penalized loss (\ref{eq:penalized}).
The remainder of this section identifies specific iteration schemes based on this useful principle.
Additional details are available in Appendix \ref{ax:derivations}.

\subsubsection{Direct Minimization}

The directional derivative of the surrogate \eqref{eq:main-surrogate} in the direction $\bW$ is
\begin{eqnarray*}
d_{\bW} g_{\rho}(\bB \mid \bB_m)
& = & - n^{-1}\tr[(\bZ_m-\bX\bB)^{\top}\bX\bW]
+\rho\tr[(\bB-\bP_m)^{\top}\bW].
\end{eqnarray*}
At a stationary point $d_{\bW}g_\rho(\bB \mid \bB_m)$ vanishes for all $\bW$. It follows that 
\begin{eqnarray}
\bzero & = & -n^{-1}\bX^\top (\bZ_m-\bX\bB)
+\rho(\bB-\bP_m) \;\: = \;\: \nabla g_\rho(\bB \mid \bB_m)
\label{eq:grad_formula}
\end{eqnarray}
and that the $\bB$ update is 
\begin{eqnarray*}
\bB_{m+1} & = & 
(n^{-1}\bX^\top\bX+\rho \bI_p)^{-1}
(n^{-1}\bX^{\top}\bZ_m+\rho \bP_m).
\end{eqnarray*}
To fit $\bB$ across a spectrum of feature subsets and values of $\rho$, we extract the thin singular value decomposition (SVD) $\bX = \bU \bSigma \bV^{\top}$ and invoke the Woodbury matrix identity
\begin{eqnarray*}
    (\bA + \bE \bC \bF)^{-1}
    =
    \bA^{-1} - \bA^{-1} \bE (\bC^{-1} + \bF \bA^{-1} \bE)^{-1} \bF \bA^{-1}
\end{eqnarray*}
with $\bA=\rho \bI_p$, $\bE = \bV$, $\bF=\bV^\top$, and $\bC=n^{-1}\bSigma^2$.
The substitutions derived in Appendix \ref{ax:mmsvd} then yield the preferred MM update
\begin{eqnarray}
    \label{eq:mmsvd}
    \bB_{m+1}
& = &     \bP_{m} + \bV 
    n^{-1} (n^{-1} \bSigma^{2} + \rho \bI_r)^{-1} 
     \left[\bSigma \bU^{\top} \bZ_{m} - \bSigma^2\bV^{\top} \bP_{m}\right]
\end{eqnarray}
involving the inverse of an $r \times r$ diagonal matrix, where $r$ equals the rank of $\bSigma$.
While the required thin SVD is expensive to extract in high dimensions, it is done only once.
This strategy successfully decouples the annealing parameter $\rho$ and model size $k$ from the multitude of matrix operations. 

\subsubsection{Steepest Descent}

Alternatively, one can take advantage of the quadratic form of \eqref{eq:main-surrogate} to derive a step size $\gamma_{m}$ in gradient descent update $\bB_{m+1} = \bB_{m} - \gamma_{m} \nabla g_{\rho}(\bB_{m} \mid \bB_{m})$.
The Taylor expansion
\begin{eqnarray*}
    g_{\rho}(\bB \mid \bB_{m})
    &=&
    g_{\rho}(\bB_{m} \mid \bB_{m})
    + \tr\left\{\nabla g_{\rho}(\bB_{m} \mid \bB_{m})^{\top}(\bB - \bB_{m})\right\} \\
    &+&
    \frac{1}{2} \tr\left\{(\bB - \bB_{m})^{\top} \nabla^{2} g_{\rho}(\bB_{m} \mid \bB_{m}) (\bB - \bB_{m})\right\},
\end{eqnarray*}
is exact.
In the particular case $\bB = \bB_{m} - \gamma \nabla g_{\rho}(\bB_{m} \mid \bB_{m})$, elementary calculus leads to the steepest descent update
\begin{eqnarray}
    \label{eq:sd}
    \gamma_{m} & = &  \frac{
        \|\bG_{m}\|_{F}^{2}
    }{n^{-1}\|\bX \bG_{m}\|_F^{2}+\rho \|\bG_{m}\|_{F}^{2}}
    \qquad
    \bB_{m+1} = \bB_{m} + \gamma_{m} \bG_{m},
\end{eqnarray}
where $\bG_{m} = \nabla g_{\rho}(\bB_{m} \mid \bB_{m})$ is defined by formula (\ref{eq:grad_formula}).

\subsection{Practical Considerations and Implementation}
We now address several issues in formulating a viable proximal distance algorithm based on our general framework.

\subsubsection{Initialization}
The penalized loss \eqref{eq:penalized} is non-convex when $S$ is a sparsity set.
Thus, solutions may be sensitive to the initial guess $\bB_{0}$.
The special cases $k=p$ and $k=0$ are natural starting points for selecting a model via cross validation. 
In these two cases the penalized loss is convex but unfortunately not strongly convex.
Choosing an initial point therefore requires some care.
We recommend minimizing the regularized loss 
\begin{eqnarray}
    \label{eq:regularized}
    h_{\lambda}(\bB) & = &  f(\bB) + \frac{\lambda}{2} \|\bB\|_{F}^{2},
    \quad \lambda \;\: > \;\: 0,
\end{eqnarray}
via the surrogate $g_{\lambda}(\bB \mid \bB_{m}) =  \frac{1}{2n} \|\bZ_{m} - \bX \bB\|_{F}^{2} + \frac{\lambda}{2} \|\bB\|_{F}^{2}$.
Strong convexity now holds, so minimization of $f_{\lambda}(\bB)$ furnishes a neutral starting point.
Any random initial value $\bB_0$ should lead to the same solution.
Note that in the special case with $k=p$ active features, the distance penalty vanishes and minimization of surrogate \eqref{eq:main-surrogate} reduces to a potentially ill-defined least squares problem.
We therefore recommend substituting the regularized surrogate \eqref{eq:regularized} for surrogate \eqref{eq:main-surrogate} to stitch the solution path from $k=p$ to $k=p-1$.

\subsubsection{Convergence Criteria}
Minimizing the penalized loss \eqref{eq:penalized} with $\rho > 0$ large generally slows down convergence of iterative methods but is required to achieve optimality \citep{beltrami1970algorithmic}.
We surmount the poor convergence characteristics of penalty methods by solving a sequence of subproblems $\eqref{eq:penalized}$ parameterized by an annealing schedule $\{\rho(t)\}_{t \ge 0}$.
The idea is to choose an annealing schedule with $\rho(0)$ small (for example, $\rho(0) = 1$) and gradually increase the penalty coefficient.
Unfortunately, the choice of annealing schedule sending $\rho \to \infty$ affects the quality of solutions.
It is therefore crucial to assess the quality of solutions in minimizing \eqref{eq:penalized} for a particular value $\rho(t)$ in the annealing schedule, the \textit{inner iterations}, and in minimizing across all subproblems, the \textit{outer iterations}.
These issues merit a brief discussion of convergence criteria.

Our previous work relied on the condition 
\begin{eqnarray*}
    |f_{\rho}(\bB_{m+1}) - f_{\rho}(\bB_{m})| & \le & \delta_f
    [1 + f_{\rho}(\bB_{m})],
\end{eqnarray*}
where $\delta_f$ is a control parameter that sets a minimum level of progress per inner iteration on a relative scale \citep{keys2019proximala,landeros2022extensions}.
In retrospect, the gradient condition
\begin{eqnarray*}
    \|\nabla f_{\rho}(\bB_{m})\| & \le &  \delta_g,
\end{eqnarray*}
is probably a safer criterion because convergence can be quite slow for large $\rho$ even under Nesterov acceleration.
The tangency condition of the MM principle conveniently allows one to compute the gradient $\nabla f_{\rho}(\bB_{m})$ through the equivalent expression $\nabla g_{\rho}(\bB_{m} \mid \bB_{m})$.
Controlling the distance term across outer iterations $t$ yields an approximate solution $\bB_{t} = \argmin~f_{\rho(t)}(\bB)$.
Thus, we impose conditions on the distance penalty
\begin{eqnarray*}
    q_{t} \le \delta_{d}
    \quad \mathrm{or} \quad
    |q_{t} - q_{t-1}| \le \delta_{q},
\end{eqnarray*}
where $q_{t} \equiv \dist(\bB_{t}, S_{k}^{\mathrm{row}})$ quantifies distance to a sparsity set.
Here the control parameter $\delta_{d}$ defines a satisfactory level of closeness to $S_{k}^{\mathrm{row}}$, and the dual criteria involving $\delta_{q}$ avoid excessive computing if progress slows dramatically. Our choices allow one to gradually guide solution estimates towards a constraint set while avoiding slow convergence due to large values of $\rho$.

\subsubsection{Proximal Distance Iteration}

Algorithm \ref{alg:pdi} summarizes the general minimization procedure in proximal distance iteration, including an early exit condition on the annealing path and Nesterov acceleration.
Warm starts are implicit in Algorithm \ref{alg:pdi} as one moves from a $\rho(t)$-penalized subproblem to the next subproblem.
The final projection step is justified when an approximate solution $\bB$ is close to a sparsity set $S_{k}^{\mathrm{row}}$ in Euclidean distance; that is, when feature selection finally stabilizes.
\begin{algorithm}[tbp]
\caption{Proximal Distance Iteration and its Convergence}
\label{alg:pdi}
\begin{algorithmic}[1]
    \footnotesize
    \State Set tolerances $\delta_{d}, \delta_{q}, \delta_{g}$, and annealing schedule $\{\rho(t)\}_{t \ge 0}$.
    \State Set target model size $k$ and sparsity level $s = 1-k/p$.
    \State Set maximum outer iterations $i_{\mathrm{outer}}$ and maximum inner iterations $i_{\mathrm{inner}}$.
    \State Set threshold for Nesterov acceleration, $i_{\mathrm{Nesterov}}$.
    \State Initialize $\bB$, and $q_{0} \gets \dist(\bB, S_{k}^{\mathrm{row}})$.
    \State Initialize counter for Nesterov acceleration $i \gets 1$.
    \For {outer iterations $t = 1,2,\ldots, i_{\mathrm{outer}}$}
        \State Initialize $\bB_{1} \gets \bB$ and $\bN_{1} \gets \bB$.
        \For {inner iterations $m = 1,2,\ldots, i_{\mathrm{inner}}$}
            \If {$\|\nabla f_{\rho(t)}(\bB_{m})\| < \delta_{g}$}
            \Comment Convergence check for fixed $\rho \equiv \rho(t)$.
                \State Break inner loop.
            \Else
                \State Solve the subproblem $\bB_{m+1} \gets \argmin g_{\rho(t)}(\bB \mid \bN_{m})$.
                \Comment e.g.~using algorithm map \eqref{eq:mmsvd} or \eqref{eq:sd}.
                \If {$f_{\rho(t)}(\bB_{m+1}) < f_{\rho(t)}(\bB_{m})$ AND $m \ge i_{\mathrm{Nesterov}}$}
                \Comment Stabililty check in Nesterov acceleration.
                    \State Accelerate $\bN_{m+1}\gets \bB_{m+1} + \frac{i-1}{i+2} (\bB_{m+1} - \bB_{m})$.
                    \State Increment $i \gets i+1$.
                \Else
                    \State Reset Nesterov acceleration; $i \gets 1$ and $\bN_{m+1} \gets \bB_{m+1}$.
                \EndIf
            \EndIf
        \EndFor

        \State Update $\bB \gets \bB_{m}$ and set $q_{t} \gets \dist(\bB, S_{k}^{\mathrm{row}})$.
        \If {$q_{t} < \delta_{d}$ OR $|q_{t} - q_{t-1}| < \delta_{q}[1 + q_{t-1}]$}
        \Comment Convergence check for penalized problem.
            \State Break outer loop.
        \EndIf
    \EndFor
    \State Project final estimate $\bB \gets P_{S_{k}^{\mathrm{row}}}(\bB)$.
\end{algorithmic}
\end{algorithm}

Let us now briefly address the theoretical convergence properties of Algorithm \ref{alg:pdi}.
The sparsity constraints described in Section \ref{sec:feature} are defined by closed but nonconvex sets. 
In principle, projection operators onto nonconvex can be multivalued.
For example, projection of the point $(5, 2, 2, 3)$ onto the sparsity set $S_{3}$ is nonunique due to ties.
Fortunately, the theoretical results from \cite{keys2019proximala} buttress Algorithm \ref{alg:pdi} under Zangwill's global convergence theorem when the penalty coefficient $\rho$ is fixed \citep{luenberger1984linear}.
First, $S_{k}^{\mathrm{row}}$ is closed for any choice of $k$ and nonempty.
Second, the $\epsilon$-insensitive loss $f(\bB)$ is coercive whenever the predictor matrix $\bX$ has full rank.
This implies the penalized loss $f_{\rho}(\bB)$ is also coercive.
Finally, observe that the surrogate generated by distance majorization is related to the possibly multivalued proximal mapping \citep{parikh2014proximal} by
\begin{eqnarray*}
    &  &\bigcup_{\bP_{m} \in P_{S_{k}^{\mathrm{row}}}(\bB_{m})}~\argmin_{\bB} \left[f(\bB) + \frac{\rho}{2}\|\bB - \bP_{m}\|^{2}\right]
    = 
    \prox_{\rho^{-1}f}\left[P_{S_{k}^{\mathrm{row}}}(\bB_{m})\right].
\end{eqnarray*}
Because set projections preserve closedness, it suffices to show that the projection operator $P_{S_{k}^{\mathrm{row}}}(\bB)$ is single-valued almost everywhere to satisfy the hypothesis of Proposition 8 from \cite{keys2019proximala}.
For fixed $\rho$ and $k$ this guarantees that all limit points of the proximal distance iterates are stationary points, $\{\bB : \nabla f_{\rho}(\bB) = \bzero \}$.
In fact, convexity of the loss $f(\bB)$ guarantees that every surrogate $g_{\rho}(\bB\mid\bB_{m})$ is $\rho$-strongly convex.
The stronger conclusion of Proposition 11 from \cite{keys2019proximala} applies and allays concerns regarding selection of projected points when $P_{S_{k}^{\mathrm{row}}}(\bB)$ is set-valued.

\subsubsection{Tuning the Sparsity Level}
\label{sec:cv}

Because a causal or optimal feature set is rarely known in advance, the effectiveness of VDA hinges on its ability to generate candidate feature sets and select an optimal model.
Cross validation is an obvious way to evaluate the classification error of candidate models.
Before discussing implementation specifics, we note that the sparsity set $S_{k}^{\mathrm{row}}$ functions as a discrete hyperparameter determining model selection.
Changing the size of the target feature set $k$ fundamentally changes the nature of the optimization problem.
In contrast, perturbing the penalty constant in a lasso penalized model merely shifts or perturbs parameter estimates along a solution path \citep{hastie2001elements}.
Model size is only indirectly determined by the penalty constant.
The main benefits of the sparsity set approach are that it (a) allows us to formulate feature selection problems directly in terms of model size, (b) relies on simple projections rather than complex combinatorial searches to guide solutions towards an optimal or nearly optimal sparsity pattern, and (c) ultimately delivers models with excellent statistical properties.

For a given dataset $\{(\by_{i}, \bx_{i})\}_{i=1}^{n}$, we implement cross validation by partitioning the samples into two subsets, one for $F$-fold cross validation and one for evaluating out-of-sample prediction.
The former subset is further split into a \textit{training set}, used to estimate a solution path spanning several model sizes $k$, and a \textit{validation set}, used to identify an optimal model size. 
Samples in the latter \textit{testing set} never appear in the cross validation procedure.
Given any set $\mathcal{T}$ of size $|\mathcal{T}|$ and model size $k$, let
\begin{eqnarray*}
    \mathrm{CV}_{\mathcal{T}}(\bB, k) & = & \frac{1}{F |\mathcal{T}|}\sum_{f=1}^F \sum_{(\by, \bx) \in \mathcal{T}} 1\{\by \neq \hat{\by}_f\}
\end{eqnarray*}
denote the classification error, on a $[0,1]$ scale, averaged over $F$ folds.
Here the $\hat{\by}_f$ are the predicted values under the parameter matrix $\bB$.
This metric allows us to choose an optimal $k$. The same random folds apply to all values of $k$.
For each $k$ we standardize features, find the best fitting $\bB$ on the training data, and evaluate the cross validation error on the training, validation, and testing sets.
The optimal $k$ over a grid is chosen to minimize validation error over a predetermined grid $\mathcal{G}$ of plausible model sizes.
Because estimates of cross validation error and therefore the selected model are sensitive to the choice of training and validation sets, we repeat the cross validation procedure several times by shuffling samples in the cross validation subset.
These replicates allow us to investigate the stability of the solution paths and the optimal choice of $k$.
Algorithm \ref{alg:cv} summarizes the entire process.
\begin{algorithm}[tbp]
\caption{Cross Validation for Sparse VDA Algorithms}
\label{alg:cv}
\begin{algorithmic}[1]
    \footnotesize
    \State {Split dataset into \textit{testing} and \textit{CV} subsets.}
    \For {each replicate $r \gets 1,R$}
        \State {Shuffle \textit{CV} subset.}
        \For {each $\mathrm{fold} \gets 1,F$}
            \State {Split \textit{CV} subset into \textit{training} and \textit{validation} subsets.}
            \State {Estimate means $\bmu$ and variances $\bsigma$ for features using the \textit{training} subset.}
            \State {Standardize all three subsets using $(\bmu, \bsigma)$.}
            \State {Fit solution path $\{\hat{\bB}_k\}_{k \in \mathcal{G}}$ to training data over a grid $\mathcal{G}$ of model sizes.}
            \State {Evaluate classification error on \textit{training}, \textit{validation}, and \textit{testing} subsets.}
        \EndFor
        \State {Estimate $\mathrm{CV}_{\mathrm{train}}$, $\mathrm{CV}_{\mathrm{val.}}$, and $\mathrm{CV}_{\mathrm{test}}$ by averaging over folds.}
        \State Rank models by the scores $\mathrm{CV}_{\mathrm{val.}}(\hat{\bB}_k, k)$.
        \State Identify and record optimal model size $k_{\mathrm{opt}}$.
    \EndFor
    \For {each metric $\gets$ ($\mathrm{CV}_{\mathrm{train}}$, $\mathrm{CV}_{\mathrm{val.}}$, $\mathrm{CV}_{\mathrm{test}}$, $k_{\mathrm{opt}}$)}
        \State Compute its median value over all $R$ replicates.
        \State Construct a 95\% credible equal-tailed interval.
    \EndFor
\end{algorithmic}
\end{algorithm}

\section{Extension to Nonlinear VDA}
\label{sec:kernelVDA}
Linear VDA for feature selection easily generalizes to nonlinear VDA based on reproducing kernels. 
Construction of decision boundaries in sparse nonlinear VDA operates by selecting representative samples called support vectors in the SVM literature \citep{wu2012nonlinear}.
Here we focus on defining and solving the optimization problem generated by nonlinear VDA.
The goal is to estimate a classification vector $\bpsi(\bx) = [\psi_{1}(\bx), \psi_{2}(\bx), \ldots, \psi_{c-1}(\bx)]$ by minimizing the objective $(2n)^{-1} \sum_{i=1}^{n} \max\{\|\by_{i} - \bpsi(\bx_{i})\| - \epsilon, 0\}^{2}$. 
Given a reproducing kernel $W(\cdot,\cdot)$, $\bpsi(\bx)$ assumes the functional form
\begin{eqnarray*}
    \psi_{j}(\bx) & = & \sum_{i=1}^{n} b_{ij} W(\bx, \bx_{i}) + b_{0j}, \qquad j=1,2,\ldots,(c-1)
\end{eqnarray*}
in predicting the vertex components from a feature vector $\bx$.
Model parameters $\bB = (b_{ij})$, including the intercepts $b_{0j}$, can be estimated by minimizing by criterion \eqref{eq:penalized} with $\bW$ substituted for $\bX$.
Nonlinear prediction is achieved through through the introduction of the reproducing kernel.
The underlying model is still linear in its parameters. 

\section{Numerical Experiments}
\label{sec:experiments}
In this section we demonstrate the feature selection capabilities of our VDA algorithms leveraging proximal distance iteration.
Our first simple example demonstrates how VDA reliably recovers sparse models without inflating (a) misclassification errors, (b) false positives in selected features, or (c) false negatives in discarded features.
Next we report benchmark results for both linear and nonlinear sparse VDA for datasets from the UCI Machine Learning Repository \citep{dua2019uci} and \textit{Elements of Statistical Learning} \citep{hastie2001elements}.
We conclude with a brief application to classical microarray expression data \citep{dettling2002supervised}.
Appendix \ref{ax:provenance} provides detailed descriptions of simulations, datasets, data preprocessing, and cross validation settings.
Appendix \ref{ax:misc} documents our choices for additional settings in VDA and describes our computing environment.

\subsection{Sparse Signal Recovery}
\label{sec:demo}

Model selection is likely to succeed only when (a) classes are sufficiently distinct, with between-class variation large compared to within-class variation, and (b) sample size is sufficiently large to overcome noise and masking.
We now demonstrate linear VDA's viability in these ideal circumstances.
Based on the simulation model described in Section \ref{ex:10clouds}), we generate synthetic datasets with $p=50$ features, $c=10$ classes, and $n=100$ or $n=1000$ samples equally distributed among classes. 
We also vary the within-class variance $\sigma^{2}$ and indirectly modulate between-class variance through an intermediate parameter $d$.
The number of causal features $p_{\mathrm{causal}} = 10$ is kept fixed throughout each scenario.
Because the simulation procedure can generate highly correlated causal features, the number $p_{\mathrm{causal}}$ should be viewed as an upper bound on estimated model size.

For a given dataset, we estimate an initial solution by minimizing the regularized model \eqref{eq:regularized} with the choice $\lambda=10^{-3}$ to avoid excessive shrinkage.
This seeds VDA with a model containing all $p$ active features.
In varying $k$ from  $p-1$ to $0$, linear VDA generates several candidate sparse solutions $\bB_{k}$ by minimizing the distance penalized objective \eqref{eq:main-surrogate}.
Thus, feature selection proceeds from a fully dense model $\bB_{p}$ to a completely sparse $\bB_{0}$.
In each case, estimating $\bB_{k-1}$ uses the previous solution $\bB_{k}$ as a warm-start.
This procedure is carried out once without cross validation to illustrate how our method generates candidate models.

\begin{figure}[bp]
    \centering
    \includegraphics[width=14.25cm,trim={0 6.5mm 0 0.5mm},clip]{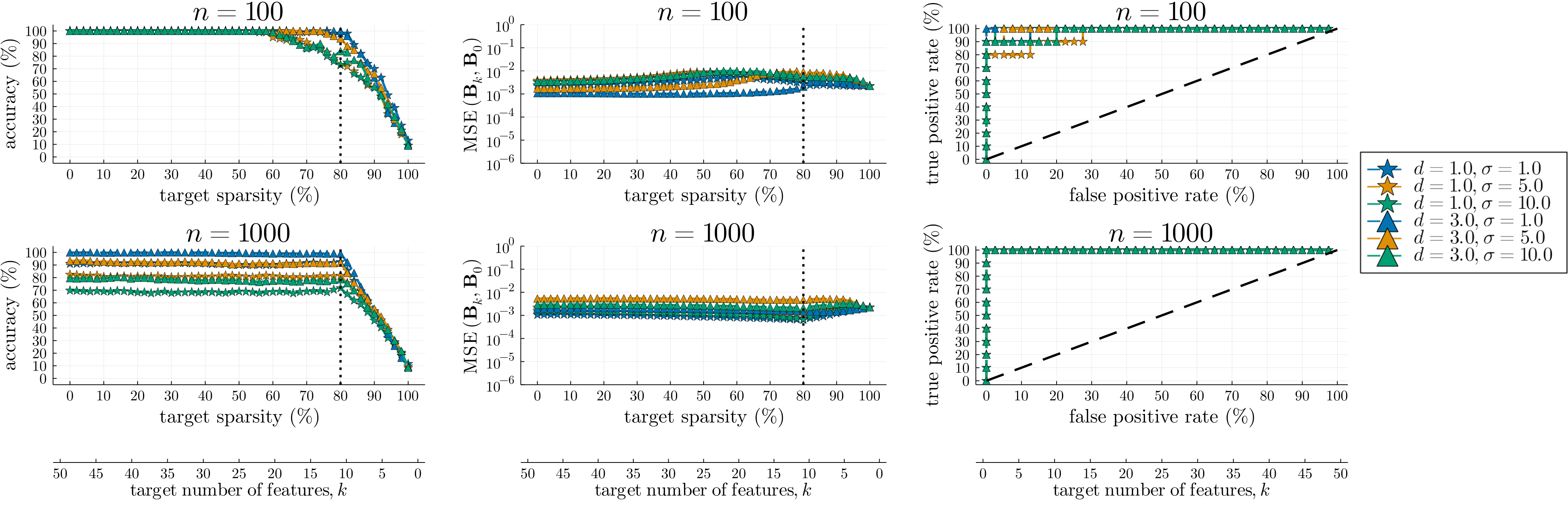}
    \caption{
        \footnotesize
        Sparse recovery under various conditions.
        The number of features, causal features, and classes are fixed to $50$, $10$, and $10$, respectively.
        Columns report classification errors, mean squared error, and ROC curves grouped by separation level $d$ (shapes; a proxy for between-class variation) and within-class variation $\sigma$ (colors).
        The true sparsity level is highlighted by a vertical line ($p_\mathrm{causal}=10$, or 80\% sparsity).
    }
    \label{fig:recovery}
\end{figure}

Our simulation study emphasizes on the following metrics:
\begin{itemize}
    \item classification accuracy on the (entire) dataset,
    \item mean squared error in model coefficients (MSE), defined on a scale relative to the ground truth $\bB_{0}$ as $\mathrm{MSE}(\bB, \bB_{0}) = p^{-1} \|\bB_{0}\|_{F}^{-2} \|\bB - \bB_{0}\|_{F}^{2}$, and
    \item the number of positive and negative calls relative to $\bB_{0}$.
\end{itemize}

Figure~\ref{fig:recovery} reports our VDA findings as receiver-operator characteristic (ROC) curves summarizing false positive and false negative rates.
It is clear that a low model size $k$ preserves a high classification accuracy (over 70\% across scenarios).
At some level the target model becomes too small for accurate classification.
The trade-off between parsimony and accuracy is reflected as an increased risk of selecting the wrong subset of features.
This in turn increases classification errors.
The green curves in Figure \ref{fig:recovery} show this tendency with $n=100$ and $\sigma=10$.
When the ratio of between-class variance to within-class variance is sufficiently high (for example, $d=3$ and $\sigma=1$), the MSE relative to $\bB_{0}$ is minimized near the true sparsity level, 80\%. 
This suggests that one can expect good estimates of model coefficients provided classes are adequately captured by a linear classifier and the dataset is sufficiently large.
The situation is only improved as sample size $n$ increases.

\subsection{Linear Classifier Benchmarks}
\label{sec:linear}

We now turn our attention to fitting linear classifiers to datasets from various applications as spelled out in Appendix \ref{ax:provenance}. On each dataset, we first fit an initial linear VDA without sparsity. We then apply $5$-fold cross validation to estimate classification errors on training, validation, and test subsets over the full range of model sizes $k \in \{1,2,\ldots,p\}$, moving from dense to sparse. The entire procedure is replicated 50 times, with each replicate shuffling the samples between training and validation subsets. In each replicate and each fold, we select an optimal model on the basis of minimum error on the validation subset and minimal model size, as described in Section \ref{sec:cv}. Finally, we summarize variability in classification errors and selected model sizes by constructing equal-tailed 95\% credible intervals from all 50 replicates. The fitted classifier includes an intercept term. The findings in Table \ref{tab:linear-results} show that our latest versionn of VDA consistently identifies a model more parsimonious than the full model.
\begin{table}[bp]
    \centering
    \caption{
        \footnotesize
        Linear classifier benchmarks. 50 replicates are summarized as 95\% equal-tailed credible intervals (minimum,median, and maximum).  Classification errors correspond to the optimal sparsity level, or, equivalently, model size, on the basis of minimal validation error.
    }
    \label{tab:linear-results}
    \tiny
    \begin{tabular}{rrrrrrrrrrrrr}
        \toprule
        & \multicolumn{3}{c}{Sparsity} & \multicolumn{3}{c}{Train} & \multicolumn{3}{c}{Validation} & \multicolumn{3}{c}{Test}\\
        \cmidrule(lr){2-4} \cmidrule(lr){5-7} \cmidrule(lr){8-10} \cmidrule(lr){11-13}
        & Med. & Min. & Max. & Med. & Min. & Max. & Med. & Min. & Max. & Med. & Min. & Max. \\
        \midrule
        iris & 50.0 & 0.0 & 69.37 & 2.92 & 2.5 & 3.33 & 3.33 & 2.5 & 5.83 & 6.67 & 5.56 & 7.78\\
        breast & 22.22 & 0.0 & 44.44 & 2.38 & 2.24 & 2.52 & 2.93 & 2.42 & 3.44 & 3.8 & 3.65 & 4.09\\
        vowel & 20.0 & 0.0 & 40.0 & 45.15 & 44.08 & 45.98 & 54.93 & 53.06 & 58.1 & 52.64 & 51.96 & 53.44\\
        letter & 0.0 & 0.0 & 0.0 & 34.3 & 34.25 & 34.39 & 34.79 & 34.64 & 35.02 & 34.5 & 34.39 & 34.61\\
        zoo & 37.5 & 6.25 & 62.5 & 0.55 & 0.0 & 1.1 & 6.58 & 4.41 & 11.1 & 6.67 & 3.33 & 10.0\\
        lymphography & 38.89 & 5.56 & 65.42 & 11.43 & 8.2 & 14.76 & 25.71 & 20.0 & 32.38 & 24.81 & 19.55 & 31.26\\
        waveform & 23.81 & 0.0 & 46.55 & 10.97 & 10.22 & 11.57 & 14.93 & 12.92 & 16.53 & 16.06 & 15.39 & 16.66\\
        optdigits & 29.69 & 10.08 & 37.5 & 2.44 & 2.36 & 2.52 & 4.29 & 4.08 & 4.55 & 4.16 & 4.0 & 4.27\\
        splice & 92.22 & 83.58 & 93.76 & 5.57 & 5.22 & 5.88 & 6.0 & 5.46 & 6.5 & 6.15 & 5.82 & 6.49\\
        HAR & 45.54 & 18.67 & 61.06 & 1.22 & 1.16 & 1.29 & 2.19 & 2.0 & 2.35 & 5.47 & 5.26 & 5.67\\        
        \bottomrule
    \end{tabular}
\end{table}

\begin{figure}[tbp]
    \centering
    \includegraphics[width=7.0cm]{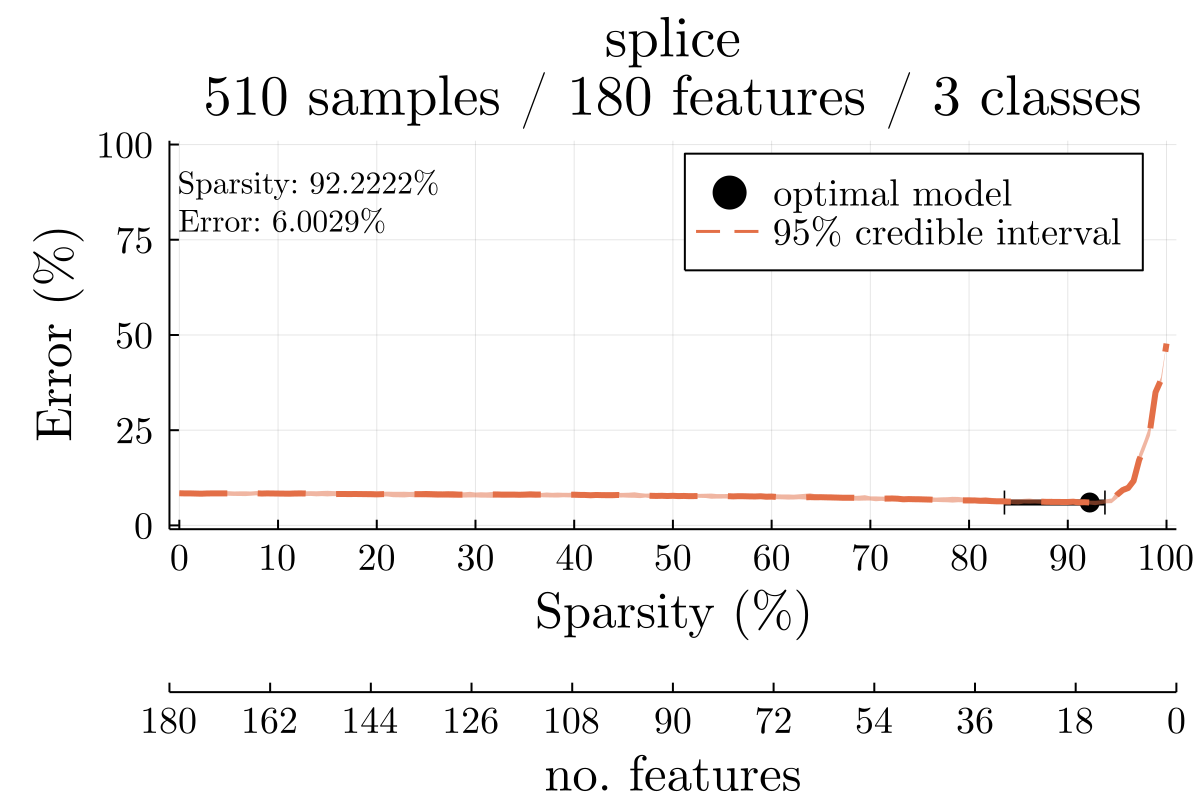}%
    \includegraphics[width=7.0cm]{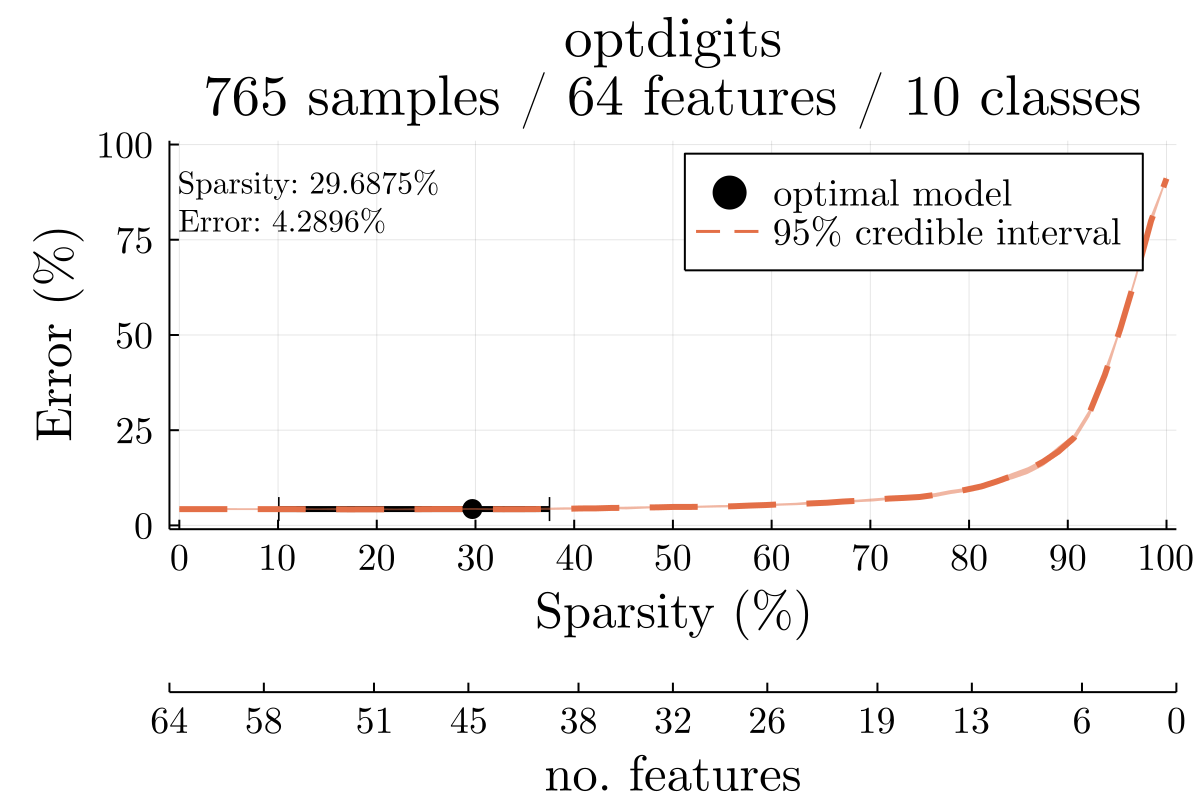}
    \caption{
        \footnotesize
        Cross validation errors on validation subsets for the \textbf{splice} and \textbf{optdigits} datasets.
        Results are based on solution path traversal from the full feature set (0\% sparsity) to the smallest possible model (100\% sparsity).
        Annotations correspond to results for the optimal model (black circle).
        The horizontal black bars highlight a credible interval for the optimal model itself to illustrate stability across replicates.
        Headings summarize the number of samples in the validation set, number of features, and number of classes.
    }
    \label{fig:linear-paths}
\end{figure}

A few of the selected examples are noteworthy. Our VDA classifier clearly fails to identify a more parsimonious model on the letter recognition example (\textbf{letters}) based on the tight intervals for classification errors. Interestingly, VDA drops at least one feature in the classic \textbf{iris} dataset, in which sepal length and sepal width fail to discriminate \textit{versicolor} and \textit{virginica}.
It is conceivable that features from the dataset \textbf{optdigits}, which summarize white pixel counts of $32 \times 32$ images along $4 \times 4$ blocks, are in fact redundant. Results on the \textbf{splice} example are also encouraging because using features closest to splice junctions is known to improve classification. In this case, the 92.22\% sparsity level corresponds to a model with 15 active features compared to 180 in the dataset. Note that each feature in this example is a binary value indicating the presence or absence of a particular nucleotide (A,G,C,T) across 60 sites. Thus, our fitted VDA classifiers reliably discriminate intron-exon, exon-intron, and non-splice sequences using approximately $15 / 3 \approx 5$ sites.

While promising, our feature selection results with VDA have an important caveat, namely stability. Figure \ref{fig:linear-paths} illustrates how classification errors vary as VDA moves towards an increasingly parsimonious model. There is a clear dip near the selected model in the \textbf{splice} example. Yet it is possible that a fitted classifier is selected because it represents the smallest model that does not increase errors compared to models earlier in the solution path.

\subsection{Nonlinear Classifier Benchmarks}
\label{sec:nonlinear}

\begin{table}[tp]
    \centering
    \caption{
        \footnotesize
        Nonlinear classifier benchmarks. 50 replicates are summarized by 95\% equal-tailed credible intervals.
    }
    \label{tab:nonlinear-results}
    \tiny
    \begin{tabular}{rrrrrrrrrrrrr}
        \toprule
        & \multicolumn{3}{c}{Sparsity} & \multicolumn{3}{c}{Train} & \multicolumn{3}{c}{Validation} & \multicolumn{3}{c}{Test}\\
        \cmidrule(lr){2-4} \cmidrule(lr){5-7} \cmidrule(lr){8-10} \cmidrule(lr){11-13}
        & Med. & Min. & Max. & Med. & Min. & Max. & Med. & Min. & Max. & Med. & Min. & Max. \\
        \midrule
        circles & 83.0 & 5.24 & 96.05 & 23.2 & 22.15 & 24.3 & 30.0 & 26.18 & 33.02 & 33.09 & 32.2 & 34.49\\
        clouds & 90.25 & 47.9 & 97.0 & 1.1 & 0.8 & 1.48 & 3.2 & 2.09 & 5.2 & 6.47 & 6.12 & 6.8\\
        vowel & 16.82 & 3.08 & 40.36 & 0.0 & 0.0 & 0.0 & 6.63 & 5.11 & 9.25 & 6.84 & 6.33 & 7.32\\
        zoo & 4.92 & 1.64 & 49.96 & 0.0 & 0.0 & 0.0 & 20.86 & 16.79 & 30.5 & 20.0 & 16.67 & 26.67\\
        waveform & 0.0 & 0.0 & 62.26 & 0.0 & 0.0 & 0.0 & 47.87 & 45.72 & 54.39 & 48.24 & 44.7 & 50.9\\
        \bottomrule
    \end{tabular}
\end{table}

We next turn to cross validation for selecting the avatars in nonlinear, kernel-based VDA. Sparse models have fewer avatars. 
Table \ref{tab:nonlinear-results} summarizes our results. Nonlinear VDA is most successful in estimating a flexible decision boundary with few avatars in the simulated datasets \textbf{clouds} and \textbf{circles}.
The reported testing error interval for \textbf{circles} compares favorably to error rates previously reported by \cite{wu2012nonlinear}, even though it is larger than the Bayes error (30\%) used in generating the data. The testing subset for dataset \textbf{vowel} is by design more difficult than the training data. Our improvement in prediction (43.07\%) over the linear VDA classifier with model selection in Table \ref{tab:linear-results} (65.26\%) is consistent with our previous findings \citep{wu2012nonlinear}. Unfortunately, our nonlinear classifiers diverge from established results on the \textbf{waveform} example, where testing errors range from 17\% to 30\% using a variety of linear and nonlinear methods.

\subsection{Cancer Benchmarks}
\label{sec:cancer}

\begin{table}[tbp]
    \centering
    \caption{
        \footnotesize
        Cancer benchmarks summarized by 95\% equal-tailed credible intervals based on 50 replicates.
    }
    \label{tab:cancer-results}
    \tiny
    \begin{tabular}{rrrrrrrrrrrrr}
        \toprule
        & \multicolumn{3}{c}{Sparsity} & \multicolumn{3}{c}{Train} & \multicolumn{3}{c}{Validation} & \multicolumn{3}{c}{Test}\\
        \cmidrule(lr){2-4} \cmidrule(lr){5-7} \cmidrule(lr){8-10} \cmidrule(lr){11-13}
        & Med. & Min. & Max. & Med. & Min. & Max. & Med. & Min. & Max. & Med. & Min. & Max. \\
        \midrule
        colon & 96.95 & 65.97 & 99.88 & 0.0 & 0.0 & 0.0 & 19.98 & 12.01 & 28.05 & 22.22 & 11.74 & 30.56\\
        srbctA & 98.77 & 91.02 & 99.61 & 0.0 & 0.0 & 0.0 & 4.04 & 0.0 & 10.14 & 5.13 & 0.0 & 14.81\\
        leukemiaA & 99.69 & 84.41 & 99.97 & 0.0 & 0.0 & 0.0 & 7.02 & 3.42 & 13.77 & 4.76 & 0.0 & 14.29\\
        lymphomaA & 99.28 & 50.84 & 99.92 & 1.49 & 0.0 & 3.03 & 9.8 & 2.5 & 17.44 & 16.67 & 6.18 & 21.6\\
        brain & 95.93 & 57.23 & 99.38 & 0.0 & 0.0 & 0.0 & 23.74 & 14.76 & 34.59 & 12.5 & 4.17 & 25.0\\
        prostate & 87.91 & 34.5 & 99.42 & 0.0 & 0.0 & 0.0 & 11.02 & 6.37 & 15.78 & 5.0 & 1.67 & 7.96\\       
        \bottomrule
    \end{tabular}
\end{table}

\begin{figure}[tbp]
    \centering
    \includegraphics[width=14.25cm]{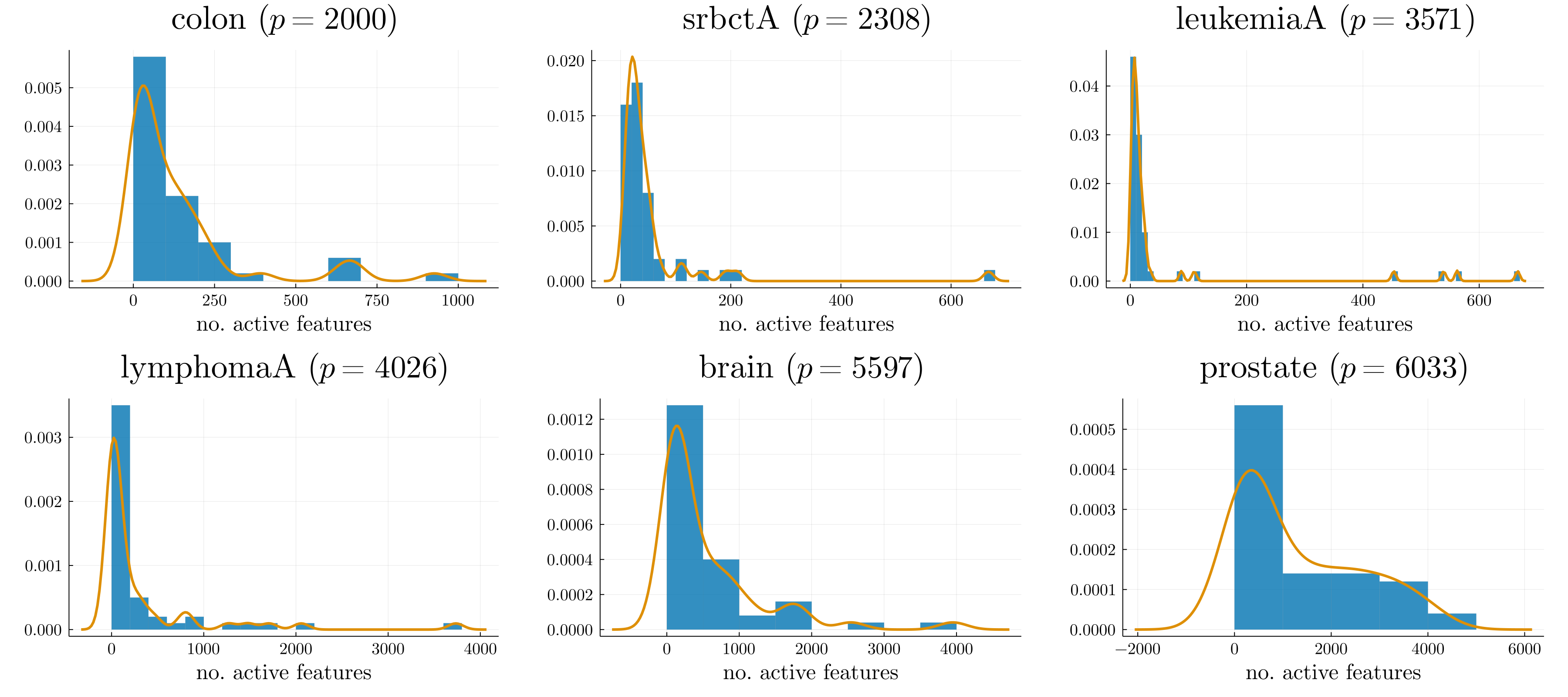}%
    \caption{
        \footnotesize
        Distribution of optimal models identified by our cross validation procedure for VDA, based on 50 replicates.
    }
    \label{fig:stability}
\end{figure}

We conclude our numerical experiments with applications to gene expression in tumors, where it is common for the number of features to dominate the number of samples, $p \gg n$. Here Table~\ref{tab:cancer-results} reports results for linear VDA classifiers across 6 datasets for different cancers as prepared by \cite{dettling2002supervised}. The \textbf{leukemia}, \textbf{prostate}, and \textbf{colon} datasets involve binary classification while the rest involve multiple classes. Our errors compare quite favorably to previous results for VDA \citep{wu2010multicategory} and hierarchical clustering \citep{dettling2002supervised}. Although our cross validation errors are not directly comparable to the best leave-one-out errors noted in Table 7 of \cite{dettling2002supervised}, we succeed in selecting models with fewer active genes then reported in Table 9. The left-skewed distributions for optimal sparsity (model size) in Figure \ref{fig:stability} support this contention in spite of the wide ranges reported for \textbf{brain} and \textbf{prostate}.

\section{Discussion and Concluding Remarks}
\label{sec:coda}

The novelty of the current version of VDA stems from the application of $\epsilon$-insensitive loss, a new majorization, distance penalties, and projection onto sparsity sets.
Once we  majorize the loss, our algorithm reduces to the classical proximal update 
\begin{eqnarray*}
    \prox_{\rho^{-1}f}(\bP_{m}) & = & \argmin_{\bB} \left[f(\bB) + \frac{\rho}{2}\|\bB - \bP_{m}\|^{2}\right];
    \quad \bP_{m} \in P_{S_{k}^{\mathrm{row}}}(\bB_{m}).
\end{eqnarray*}
Computing this proximal update involves solving the nonlinear equations
\begin{eqnarray*}
    (\bX^{\top}\bW_{\bB}\bX + \rho \bI) \bB = \bX^{\top} \bW_{\bB} \bY + \rho \bP_{m},
\end{eqnarray*}
where $\bW_{\bB}$ is a diagonal matrix depending on the residuals $\by_{i} - \bB^{\top} \bx_{i}$.
This structure suggests the standard iteratively reweighted least squares (IRLS) strategy noticed by other authors in robust classification \citep{groenen2008svmmaj,burg2016gensvm,nguyen2017iterativelyreweighted}.
Sequential minimization via our surrogate \eqref{eq:main-surrogate} is effectively an IRLS algorithm for fitting VDA classifiers. 

Our work has its limitations.
While our empirical results are satisfying, we have not characterized the statistical properties of our estimators. Notably, we still lack conditions implying the consistency of VDA  \citep{lange2008mm,wu2010multicategory}.
The simulation study in Section \ref{sec:demo} makes the case that our model estimates successfully reduce mean squared error (Figure~\ref{fig:recovery}).
Statistical theory would be especially helpful in understanding the stability of model selection under cross validation.
Figure~\ref{fig:stability} suggests that VDA is strongly biased towards selecting parsimonious models on datasets believed to have a small number of informative features.
Unfortunately, cross validation replicates often yield candidate models with different numbers of active features.
Dense models can be recovered, as illustrated by the prostate example summarized in  Figure~\ref{fig:stability}.
Another limitation is that only one or a few driving features may be selected out of a cluster of highly correlated features.
This tendency may be beneficial in finding causal features in biological models.
Finally, our code implementation only relies on BLAS calls to exploit parallelism in linear algebra operations.
Our algorithms may have additional parallel structure gone unnoticed.

Our VDA benchmarks do not report computational times required for solving a single penalized problem or running cross validation.
Without supporting theory, it seems premature to pit VDA against the wide array of readily available software for classification.
Anecdotally, cumulative  VDA fitting times (across every fold, model size, and cross validation replicate) range from about 5 seconds for the iris, zoo, lymphography data to several minutes for the splice, letters, colon, and lymphoma data.
The most time consuming dataset to fit was the HAR example, which required $>7$ hours to generate 50 cross validation replicates with 5 folds each, or about $8.4$ minutes per replicate on average.
These observations are encouraging, particularly in the more interesting case of sparse linear VDA.
While our model selection technique permits flexibility in nonlinear decision boundaries, we find that fitting nonlinear VDA classifiers is noticeably slower than linear VDA.
The additional computational cost and loss of interpretability in nonlinear VDA is unjustified in cases where a linear classifier is strictly better (see for example the zoo benchmarks in Tables \ref{tab:linear-results} and \ref{tab:nonlinear-results}).
The SVD-based variant of our software slows noticeably with a large number of samples.
Additional computational tricks, beyond parallel linear algebra operations, are needed to scale our VDA algorithms to datasets with sample sizes $n > 10^{5}$.
Mature classification packages such as LIBSVM \citep{chang2011libsvm} may provide insight in overcoming computing limitations.

Repeated cross validation is not the only technique for evaluating candidate models.
For instance, the method of \textit{stability selection} described by \cite{meinshausen2010stability} is attractive for its theoretical finite-sample guarantees in controlling family-wise errors.
Alternatively, \textit{selecting features in groups} can help overcome multicollinearity \citep{alexander2011stability}.
Ideally group selection should also control candidate model sizes.
Without going into details, it is possible to project onto sets that limit the number of groups selected and the number of features selected per group.
The downside of doubly sparse cross validation is that involves searching over a 2-dimensional tuning-constant grid.
The \textit{percentile lasso} \citep{roberts2014stabilizing} is another attractive device for stabilizing model selection.
Our repeated cross validation procedure, similar to the percentile lasso, attempts to control variability by using the median of optimal model sizes across replicates.
The distributions reported in Figure~\ref{fig:stability} suggest that adapting feature selection to account for the variety of models generated by replicate cross validations may, in fact, improve credible intervals.
Incorporating a Bayesian perspective into sparse VDA might be based on the the peaks in Figure~\ref{fig:stability}, but again model selection is apt to become even more time consuming.

In conclusion, we have demonstrated both the prediction accuracy and the feature selection capabilities of our modified VDA algorithms across a range of simulated and real datasets.
Our numerical examples confirm the promise of incorporating feature selection into VDA with distance to constraint-set penalties.
Combining our estimation procedures with cross validation overcomes the need to specify a particular model size.
Replicating cross validation further improves the reliability of feature selection with sparse VDA and provides a straightforward mechanism for honest reporting.
In the case of kernel VDA, our proposed methods allow for the simultaneous selection of the support vectors defining the decision boundaries separating multiple classes.
Our future work on VDA will seek to address gaps in theory, the caveats of repeated cross validation, and the extension of VDA to clustering problems in analogy with SVM-based clustering.

\bibliographystyle{chicago}
\bibliography{references}

\newpage

\begin{appendices}
\section{Derivations}
\label{ax:derivations}

\subsection{Directly minimizing surrogate \eqref{eq:main-surrogate}}
\label{ax:mmsvd}
According to the orthogonality relation $\bV^\top\bV=\bI_r$ and Woodbury's formula, the required inverse is
\begin{eqnarray*}
    [n^{-1} \bV \bSigma^2 \bV^{\top} + \rho \bI_{p}]^{-1}
    &=&
    \rho^{-1} \left[
        \bI_{p} - \rho^{-1} \bV (n \bSigma^{-2} + \rho^{-1} \bV^{\top} \bV)^{-1} \bV^{\top}
    \right] \\
    &=&
    \rho^{-1} \left[
        \bI_{p} - \rho^{-1} \bV (n \bSigma^{-2} + \rho^{-1} \bI_{r})^{-1} \bV^{\top}
    \right] \\
    &=&
    \rho^{-1} \left[
        \bI_{p} - n^{-1} \bV (\rho \bI + n^{-1} \bSigma^{2})^{-1} \bSigma^2 \bV^{\top} 
    \right].
\end{eqnarray*}
Multiplying the above inverse by $\rho \bP_m$ yields
\begin{eqnarray*}
    \bP_m
    -
    n^{-1} \bV (n^{-1} \bSigma^{2} + \rho \bI_{r})^{-1} \bSigma^{2} \bV^{\top} \bP_m.
\end{eqnarray*}
Likewise multiplying the above inverse by $n^{-1} \bX^{\top} \bZ_m$ gives
\begin{eqnarray*}
    && n^{-1} \rho^{-1} \left[
        \bV \bSigma \bU^{\top}
        -
        n^{-1} \bV (n^{-1} \bSigma^{2} + \rho \bI_{r})^{-1} \bSigma^{2} \bV^{\top} \bV \bSigma \bU^{\top}
    \right] \bZ_m \\
    &=&
    n^{-1} \rho^{-1} \bV \left[
        \bI_r
        -
        n^{-1} (n^{-1} \bSigma^{2} + \rho \bI_r)^{-1} \bSigma^{2}
    \right] \bSigma \bU^{\top} \bZ_m \\
    &=&
    n^{-1} \rho^{-1} \bV (n^{-1} \bSigma^{2} + \rho \bI_r)^{-1} \left[
        (n^{-1} \bSigma^{2} + \rho \bI_r) - n^{-1} \bSigma^{2}
    \right] \bSigma \bU^{\top} \bZ_m \\
    &=&
    n^{-1} \bV (n^{-1} \bSigma^{2} + \rho \bI_r)^{-1} \bSigma \bU^{\top} \bZ_m.
\end{eqnarray*}
Summing these two results leads to the desired update
\begin{eqnarray*}
    \bB_{m+1} &=& \bP_m
    + n^{-1} \bV (n^{-1} \bSigma^{2} + \rho \bI_{r})^{-1} \left[
        \bSigma \bU^{\top} \bZ_m-\bSigma^{2} \bV^{\top} \bP_m
    \right] .
\end{eqnarray*}

\section{Dataset Descriptions and Provenance}
\label{ax:provenance}

\subsection{Preprocessing}

Throughout each example we standardize each feature vector $\bx_{i}$ to have mean $\bzero$ and unit variance by estimating sample means $\bmu = \frac{1}{n} \sum_{i=1}^{n} \bx_{i}$ and variances $\bsigma^{2} = \frac{1}{n-1} \sum_{i=1}^{n} (\bx_{i} - \bmu)^{2}$, where operations are applied element-wise.
We then apply a $Z$-score transformation $\bx \mapsto (\bx - \bmu) / \bsigma$.
Location and scale estimates are based on a \textit{training} subset only, and the $Z$-score transformation is applied to all subsets in cross validation.
In the nonlinear case, we estimate the variance in distances between dissimilar samples (that is, when true class labels are distinct) using the median and set a scale parameter $\gamma = 13/10 \times \mathrm{median}$ for radial basis functions (RBF) kernels.

For real-world datasets we take care to drop missing data.
In addition, we make the choice to drop any non-numeric features (e.g. categorical data) \textit{that are not already coded}.
We note this in each description where applicable.

\subsection{Gaussian Clouds (clouds)}
\label{ex:clouds}
As described by \cite{wu2012nonlinear}, we simulate six Gaussian clouds with centers $(x,y)$ equally spaced along the unit circle in $\Real^{2}$.
\[
    \theta = \begin{cases}
        0\, \text{or}\, \pi & \text{for class 1}, \\
        \frac{\pi}{3}\, \text{or}\, \frac{4\pi}{3} & \text{for class 2}, \\
        \frac{2\pi}{3}\, \text{or}\, \frac{5\pi}{3} & \text{for class 3},
    \end{cases}
    \quad \text{with} \quad (x,y) = (\cos(\theta), \sin(\theta)).
\]
Each class has equal probability and centers are sampled uniformly conditional on the chosen class.
Points about a given cloud are simulated with variance-covariance matrix $\sigma^{2} \bI_{2 \times 2}$.
Thus, higher variability increases class overlap and hence classification difficulty.
We set $\sigma=0.25$ and split the data into training, validation, and testing subsets of sizes 200, 50, and 1000, respectively, in 5-fold cross validation.

\subsection{Nested Circles (circles)}
\label{ex:circles}
The recipe described by \cite{wu2012nonlinear} samples two features from the circle $\{(x,y) : x^{2} + y^{2} < c\}$ and assigns them to one of $c$ classes.
The intermediate values $r = \lceil x^{2} + y^{2}\rceil$ and $r^{\prime}$ are used to simulate class assignment via the rule
\[
    \text{class} = \begin{cases}
        r & \text{with probability}\, p \\
        r^{\prime} & \text{with probability}\, 1-p.
    \end{cases}
\]
Here $\lceil \cdot \rceil$ is the ceiling function, $r^{\prime}$ is any class in $\{1,2,\ldots,c\}$ other than $r$ which is sampled uniformly, and $1-p$ is the Bayes error controlling the difficulty of classification.
Our benchmark takes $c=3$ and $p=0.8$ to simulate a dataset with $1250$ samples.
We split the data into training, validation, and testing subsets of sizes 200, 50, and 1000, respectively, in 5-fold cross validation.

\subsection{Waveform (waveform)}
\label{ex:waveform}
This dataset is based on convex combinations of triangular waveforms that was originally presented by \cite{breiman1984classification} and featured in Example 12.7.1 of \textit{The Elements of Statistical Learning} \citep{hastie2001elements}.
We simulate $p=21$ features across $c=3$ classes using the recipe
\begin{eqnarray*}
    X_{j} &= U h_{1}(j) + (1-U) h_{2}(j) + \epsilon_{j} & \text{for class 1}, \\
    X_{j} &= U h_{1}(j) + (1-U) h_{3}(j) + \epsilon_{j} & \text{for class 2}, \qquad j=1,2,\ldots,21 \\
    X_{j} &= U h_{2}(j) + (1-U) h_{3}(j) + \epsilon_{j} & \text{for class 3},
\end{eqnarray*}
where $U$ is a uniform deviate on $[0,1]$ and the $\epsilon_{j}$ are independent standard normal deviates.
The triangular waveforms $h_{1}(j) = \max\{6 - |j-11|, 0\}$, $h_{2}(j) = h_{1}(j-4)$, and $h_{3}(j) = h_{1}(j+4)$ are centered at $j=11$, $j=15$, and $j=7$, respectively.
A total of 1375 samples are simulated with training, validation, and testing subsets of sizes 300, 75, and 1000, respectively, in $5$-fold cross validation.

\subsection{Sparse Recovery Model (10clouds)}
\label{ex:10clouds}

This recipe is adapted from the \textbf{clouds} example and a recipe from \cite{wang2007l1norm} (see Section 3.1).
Given a target number of features $p$ and classes $c$, we simulate features by sampling from $\mathcal{N}(\bmu, \bSigma_{p \times p})$, where
\begin{align*}
    \theta_{\ell} &= \frac{\pi}{3} + \frac{2(\ell-1)\pi}{c}, \\
    a_{\ell} &= \begin{cases}
        \cos\theta_{\ell}, & \ell~\text{is even} \\
        \sin\theta_{\ell}, & \text{otherwise}
    \end{cases}, \\
    \mu_{j} &= \begin{cases}
        d \times \sgn(a_{\ell}), & \text{for}~j=1,2,\ldots,c \\
        0, & \text{otherwise}
    \end{cases},~\text{and} \\
    \Sigma_{ij} &= \begin{cases}
        \sigma, & i=j~\text{and}~j=1,2,\ldots,c \\
        1/\sigma, & i=j~\text{and}~j>c \\
        0, & \text{otherwise}
    \end{cases}.
\end{align*}
For each sample $\bx_{i}$ we simulate the response $\by_{i} = \bB^{\top} \bx_{i}$ where each row $\bb_{j}$ of $\bB$ is assigned
\[
    \bb_{j} = \begin{cases}
        c^{-1} \sgn(a_{\ell}) \bv_{j}, & j = 1,2,\ldots,c \\
        \bzero, & \text{otherwise}
    \end{cases}
\]
and each $\bv_{j}$ is one of $c$ vertices assigned to classes.
These choices increase the likelihood that the simulated dataset contains close to $c$ informative features.
Large values of parameter $d$ induce separation between classes whereas $\sigma$ controls variability within each class.

\subsection{UCI Datasets}
\label{ex:uci}
We select 8 datasets from the UCI Machine Learning Repository \citep{dua2019uci}, all of which are in the overdetermined regime.
Table \ref{tab:uci} summarizes the characteristics of each selected example.

\begin{table}[tbp]
    \centering
    \caption{
        \footnotesize
        Summary of datasets from UCI Machine Learning Repository and corresponding cross validation settings.
        Values in parentheses indicate the number of samples, features, or classes dropped from the original dataset.
    }
    \label{tab:uci}
    \footnotesize
    \begin{tabular}{rrrrrrr}
        \toprule
        & \# classes & \# samples & \# features & \# folds & Train / Test \\
        \midrule
        iris & 3 & 150 & 4 & 3 & 120 / 30 \\
        lymphography & 4 & 148 & 18 & 3 & 105 / 43 \\
        zoo          & 7 & 101 & 16 (1) & 3 & 91 / 10 \\
        bcw          & 2 & 699 & 9 & 5 & 562 / 137 \\
        splice       & 3 & 3186 & 180 & 5 & 2549 / 637 \\
        letters      & 26 & 20000 & 16 & 5 & 16000 / 4000 \\
        optdigits    & 10 & 5620 & 64 & 5 & 3823 / 1797 \\
        HAR          & 6 & 10299 & 561 & 5 & 7352 / 2947 \\
        \bottomrule
    \end{tabular}
\end{table}

\subsection{Cancer Datasets}
\label{ex:cancer}

Our experiments on cancer microarray expression data is based on 6 preprocessed datasets as described by \cite{dettling2002supervised} and previously available from {https://stat.ethz.ch/$\sim$dettling/supercluster.html}.
The detailed study of \cite{dettling2002supervised} reports leave-one-out cross validation error rates on leukemia (1.39\%), prostate cancer (4.90\%), colon cancer (16.13\%), Small-Blue-Round-Cell-Tumor (SBRCT) cancer (0.00\%), lymphoma (0.00\%), and brain cancer (11.90\%) using a supervised clustering method that compared favorably against existing literature at the time.
Notably, all 6 examples are in the underdetermined regime.
\begin{table}[tbp]
    \centering
    \caption{
        \footnotesize
        Summary of cancer microarray expression data and corresponding cross validation settings.
    }
    \label{tab:cancer}
    \footnotesize
    \begin{tabular}{rrrrrrr}
        \toprule
        & \# classes & \# samples & \# features & \# folds & Train / Test & References \\
        \midrule
        leukemia & 2 & 72 & 3571 & 3 & 58 / 14 & \cite{golub1999molecular} \\
        prostate & 2 & 102 & 6033 & 3 & 82 / 20 & \cite{singh2002gene} \\
        colon    & 2 & 62 & 2000 & 3 & 50 / 12 & \cite{alon1999broad} \\
        SRBCT    & 4 & 63 & 2308 & 3 & 50 / 13 & \cite{khan2001classification} \\
        lymphoma & 3 & 62 & 4026 & 3 & 50 / 12 & \cite{pomeroy2002prediction} \\
        brain    & 5 & 42 & 5597 & 3 & 34 / 8 & \cite{alizadeh2000distinct} \\
        \bottomrule
    \end{tabular}
\end{table}

\subsection{Vowel Dataset}
\label{ex:vowel}

This dataset is taken from \textit{Elements of Statistical Learning}.
It consists of $n=990$ samples with $p=10$ features and $c=11$ classes.
We use the original train / test split of 528 / 462 in 5-fold cross-validation.
A full original description of the dataset is available at \texttt{https://hastie.su.domains/ElemStatLearn/}.

\subsection{Visualizations of Simulated Examples}

\begin{figure}[tbp]
    \centering
    \includegraphics[width=4.75cm]{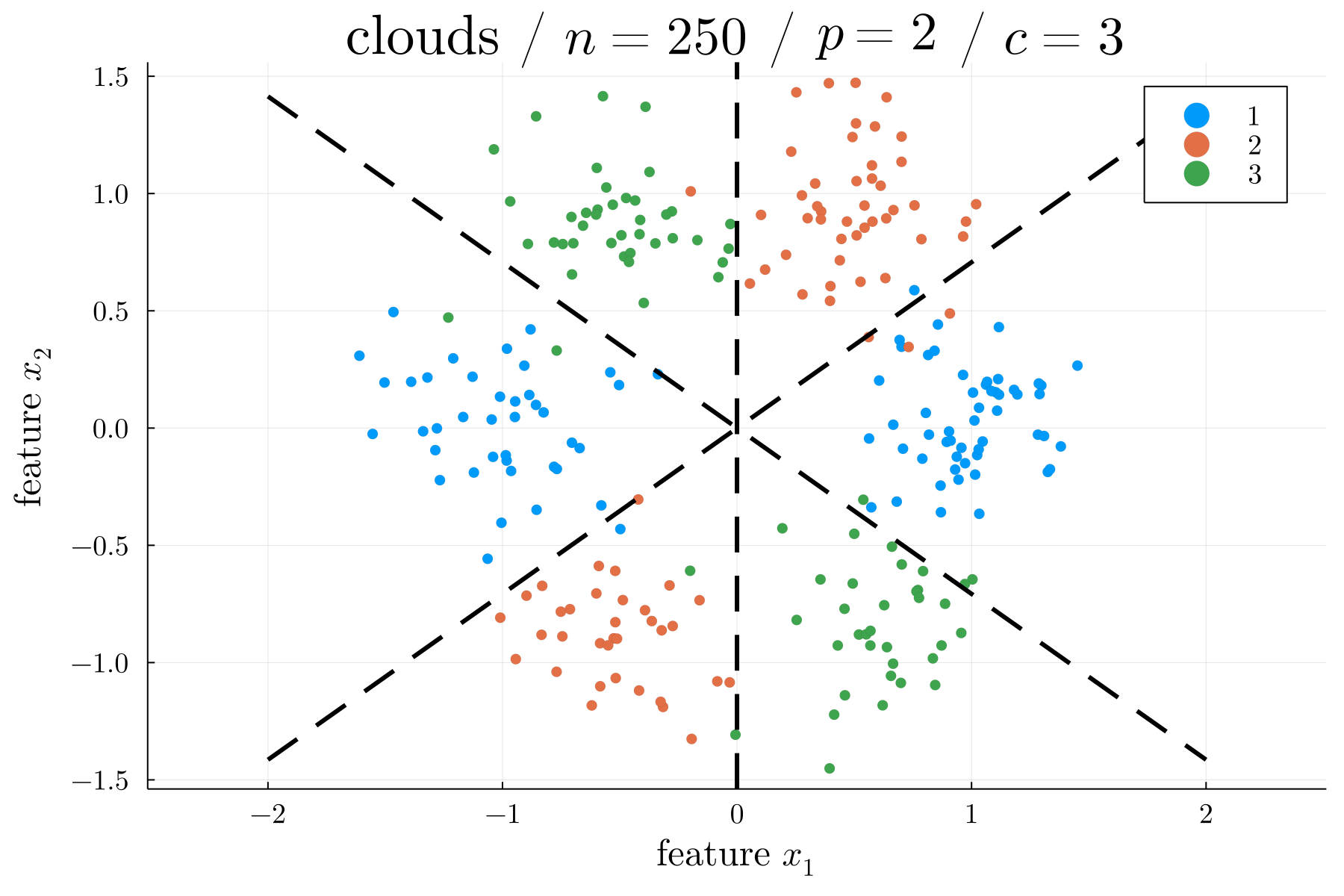}%
    \includegraphics[width=4.75cm]{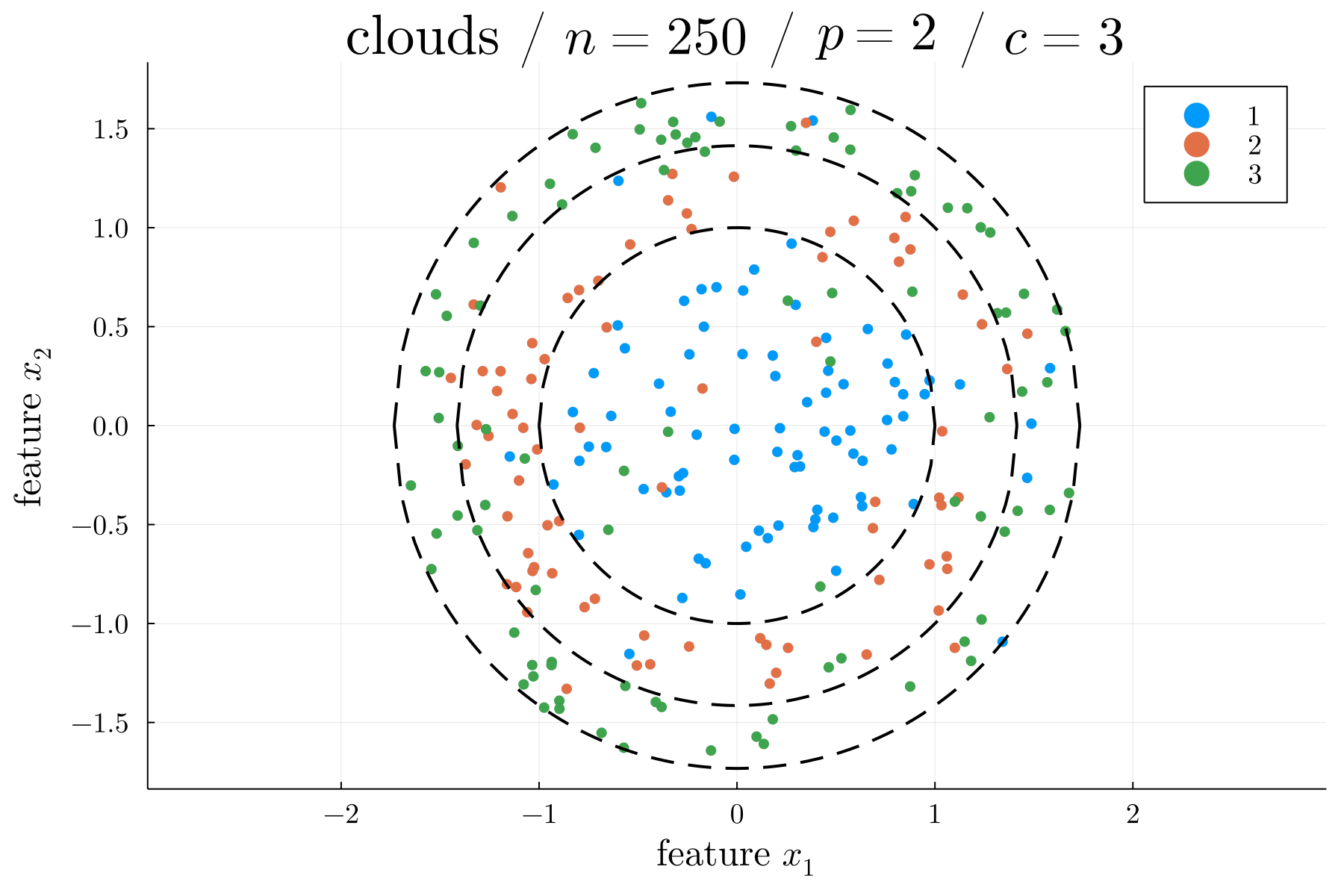}%
    \includegraphics[width=4.75cm]{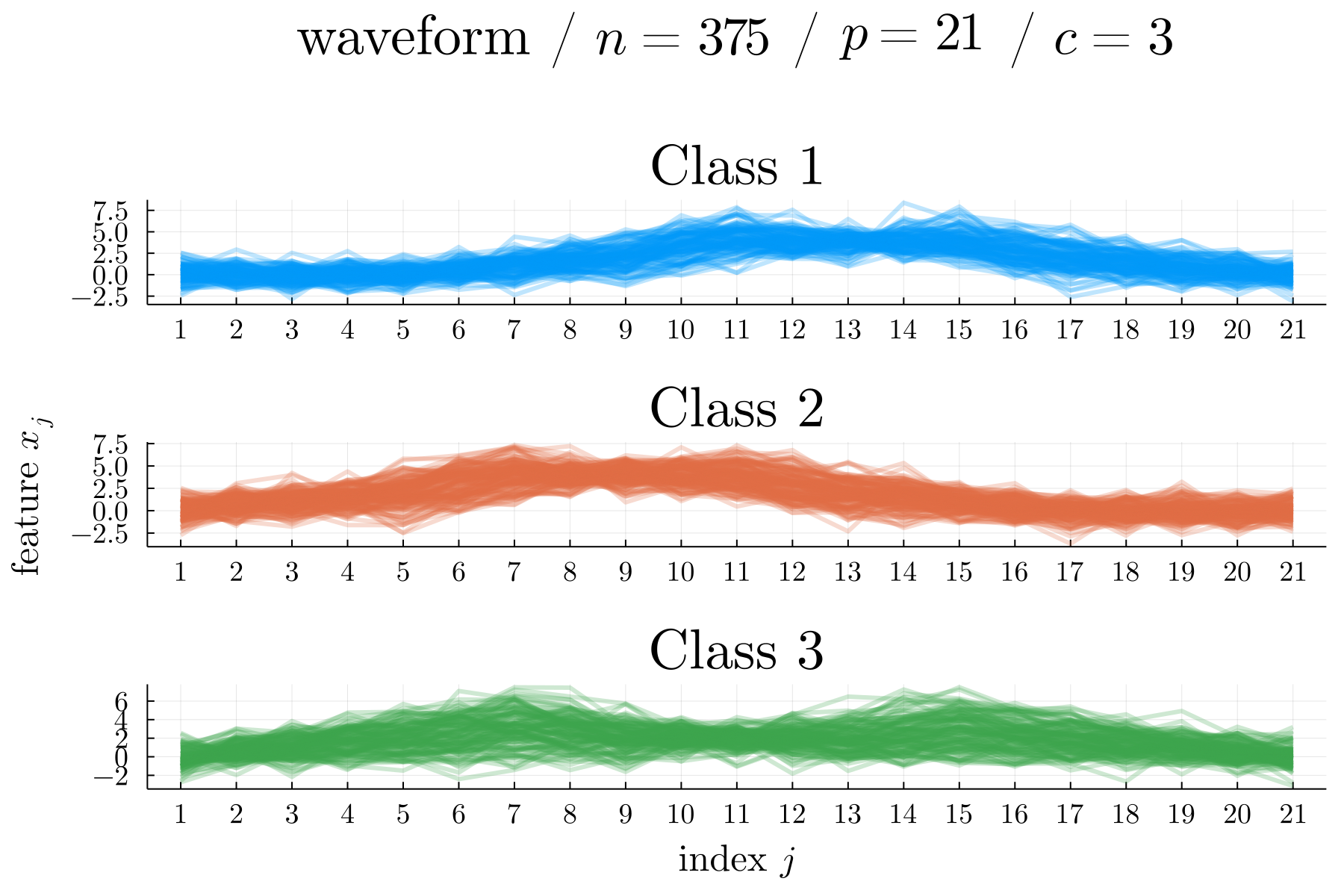}
    \includegraphics[width=4.75cm]{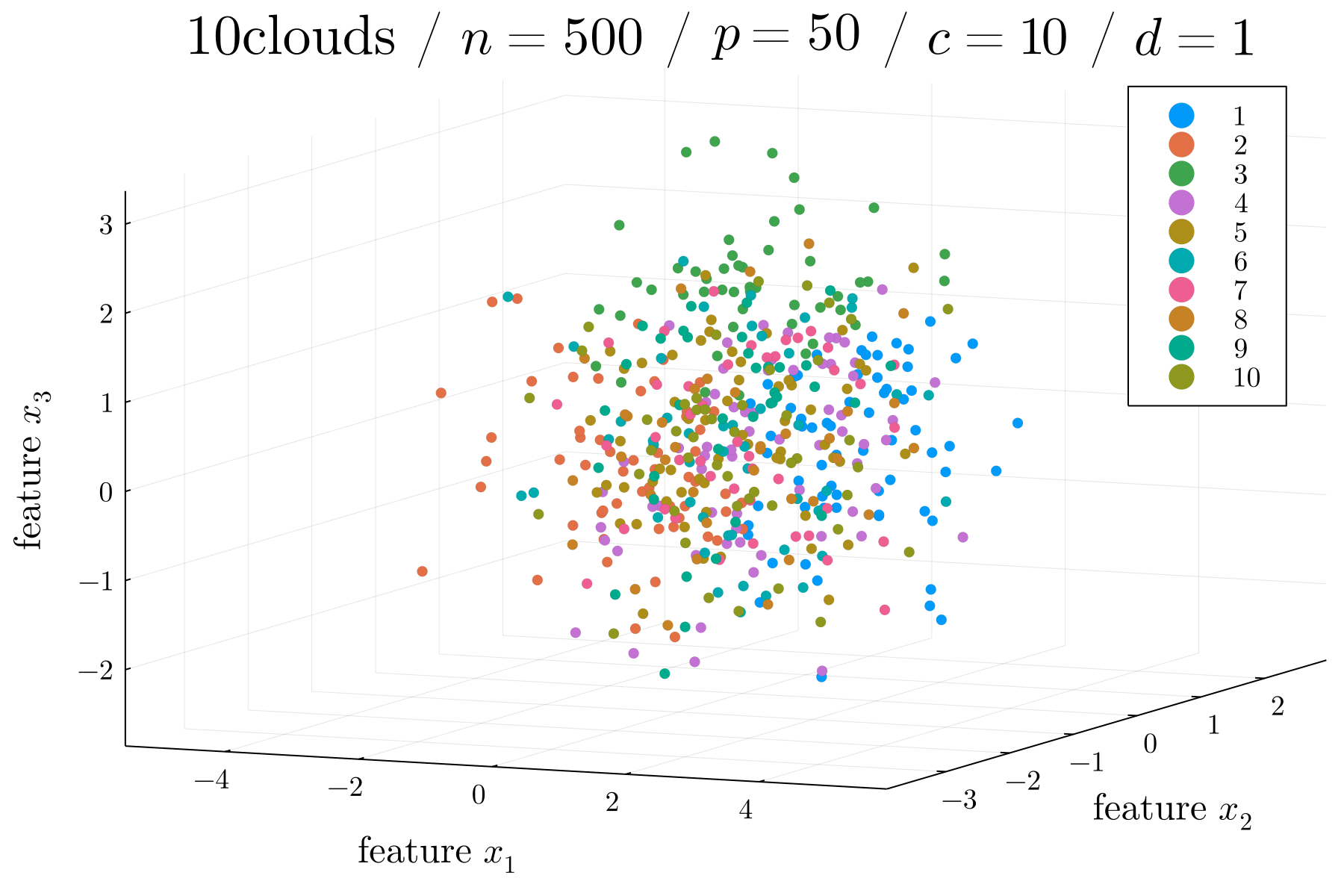}%
    \includegraphics[width=4.75cm]{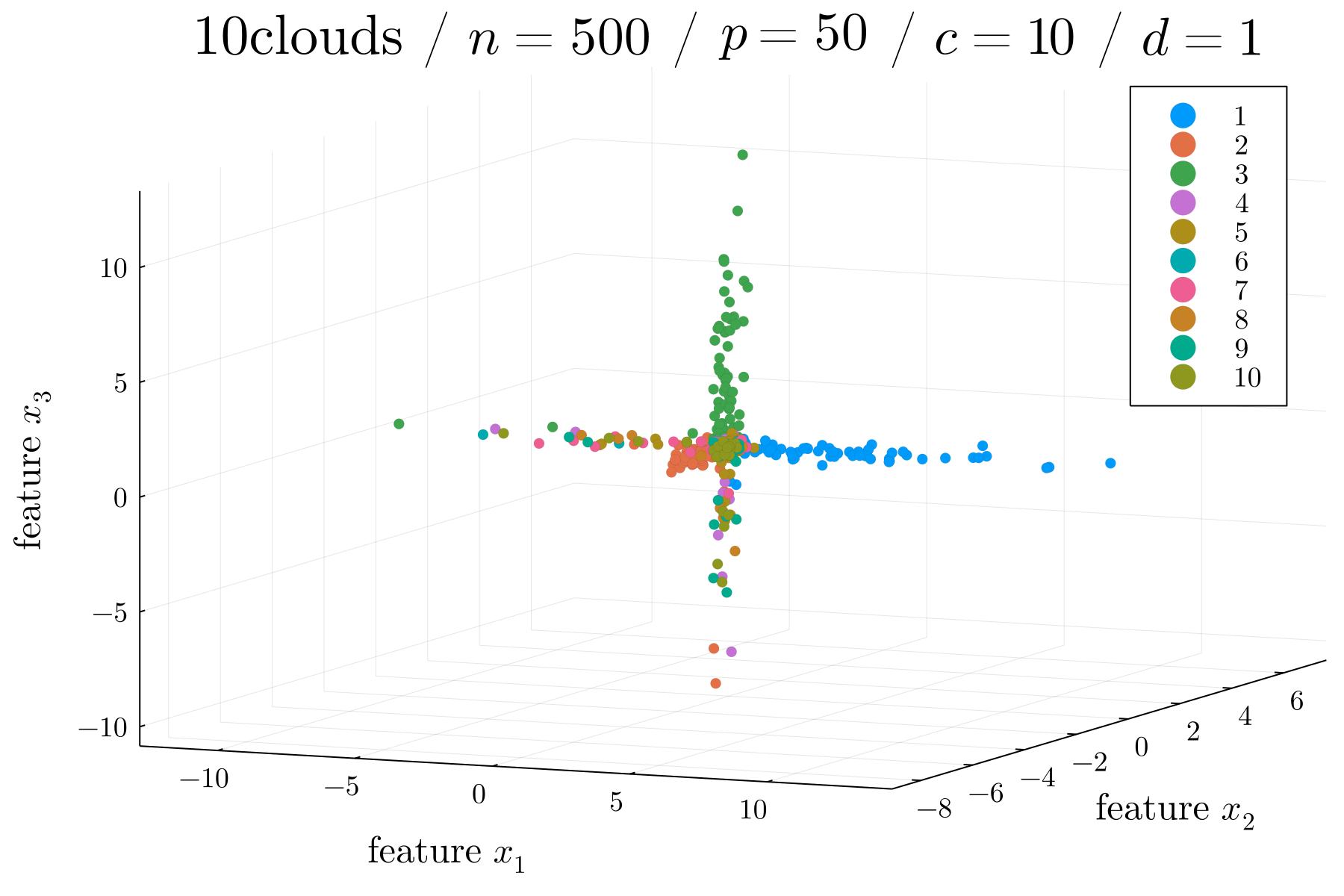}%
    \includegraphics[width=4.75cm]{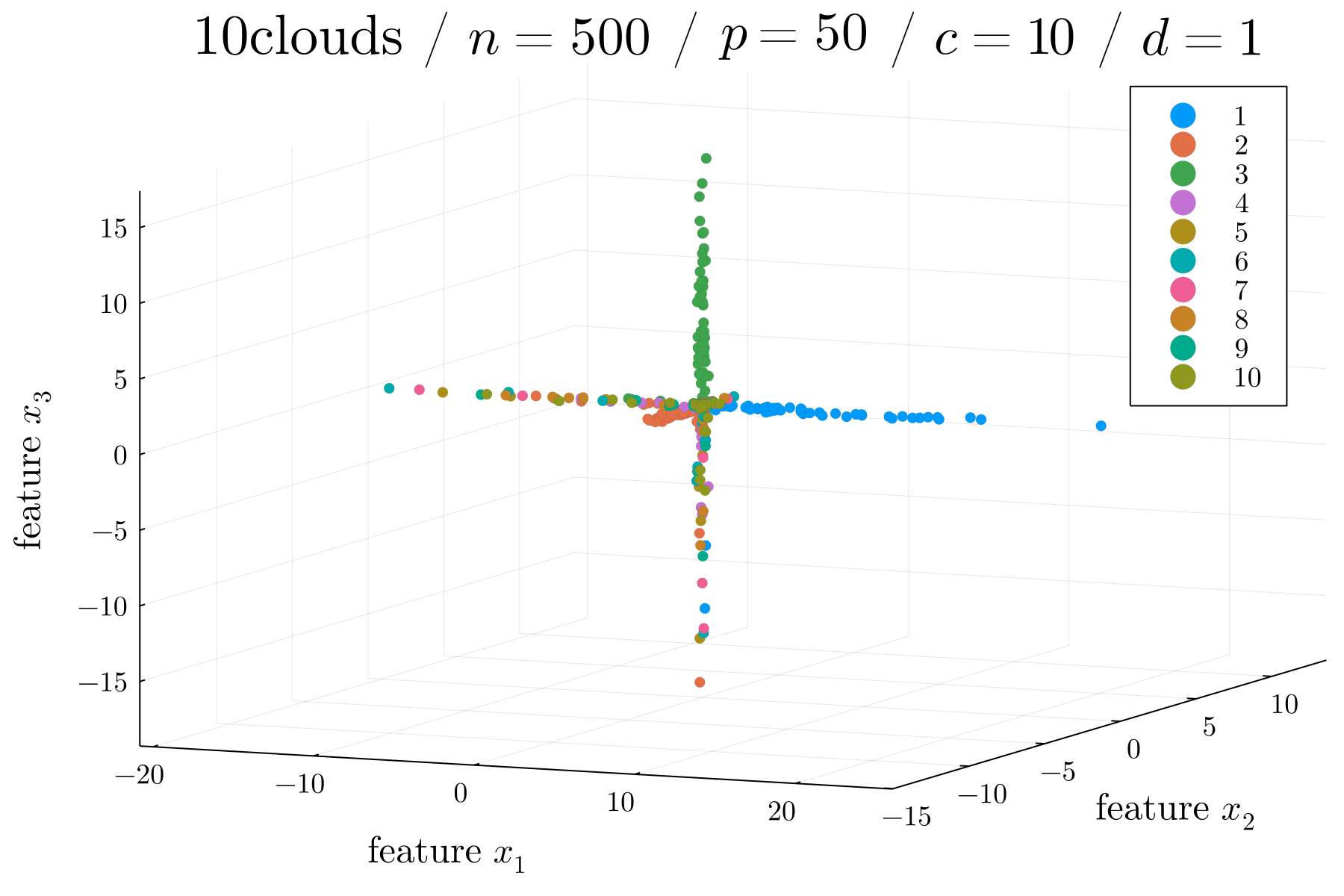}
    \includegraphics[width=4.75cm]{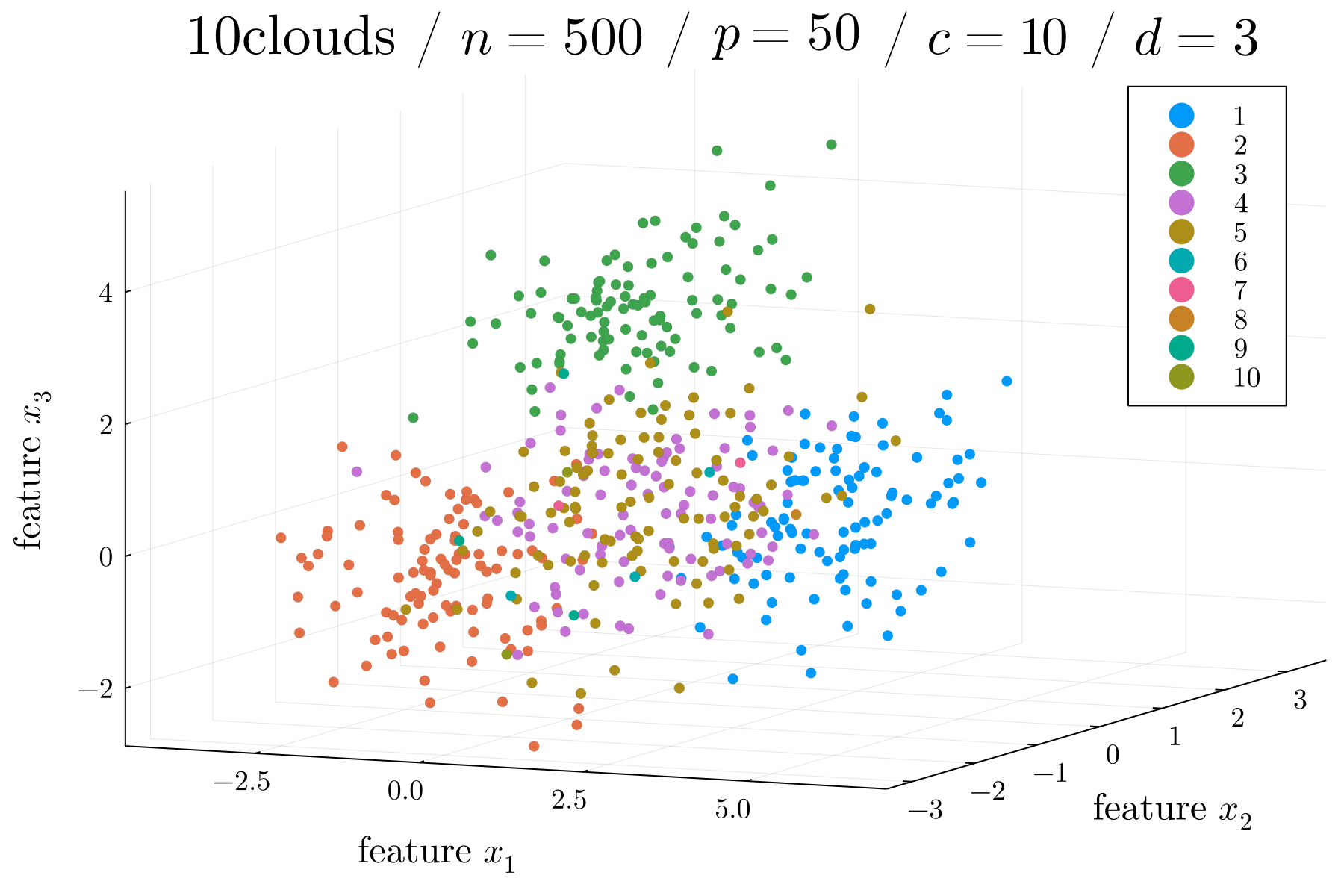}%
    \includegraphics[width=4.75cm]{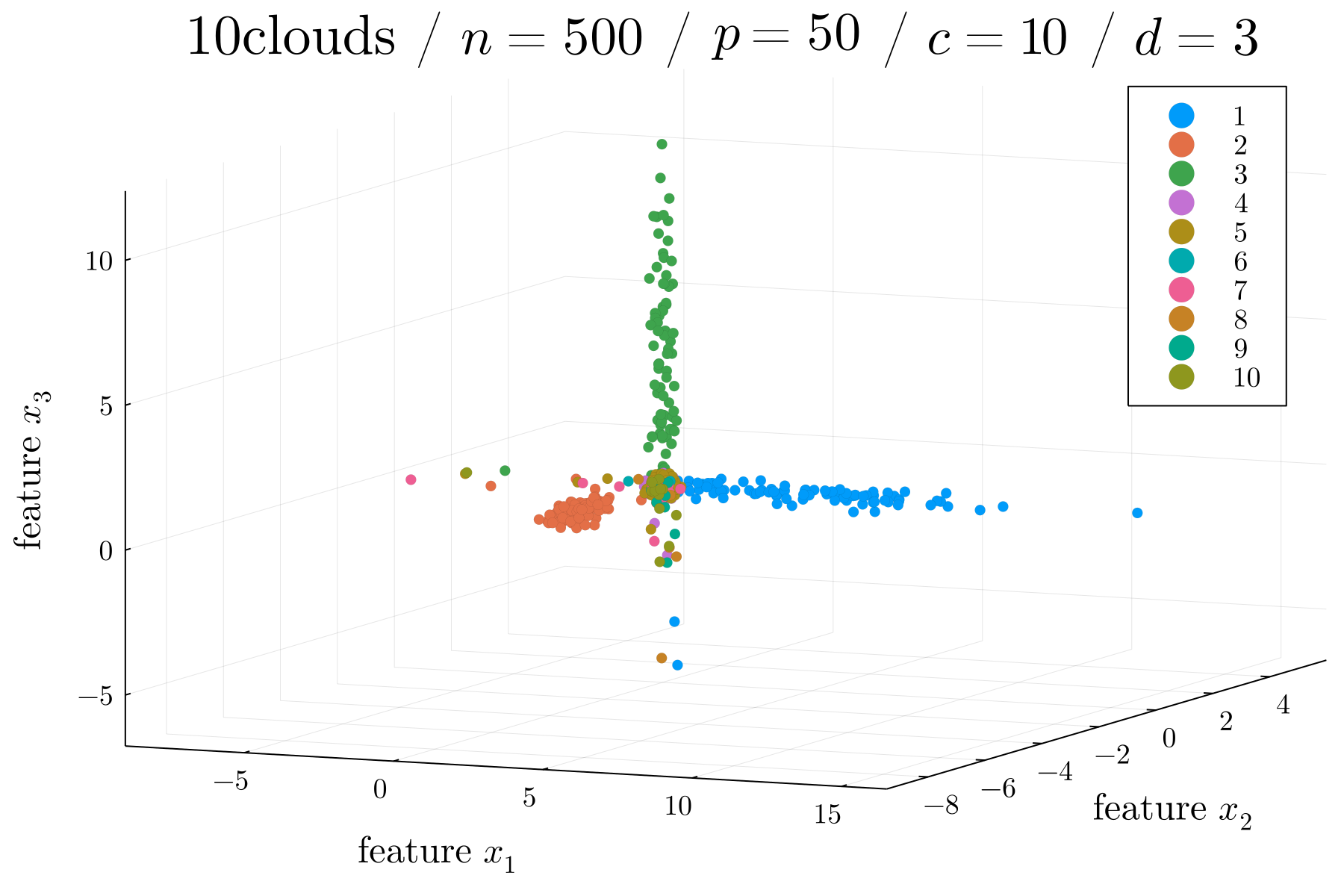}%
    \includegraphics[width=4.75cm]{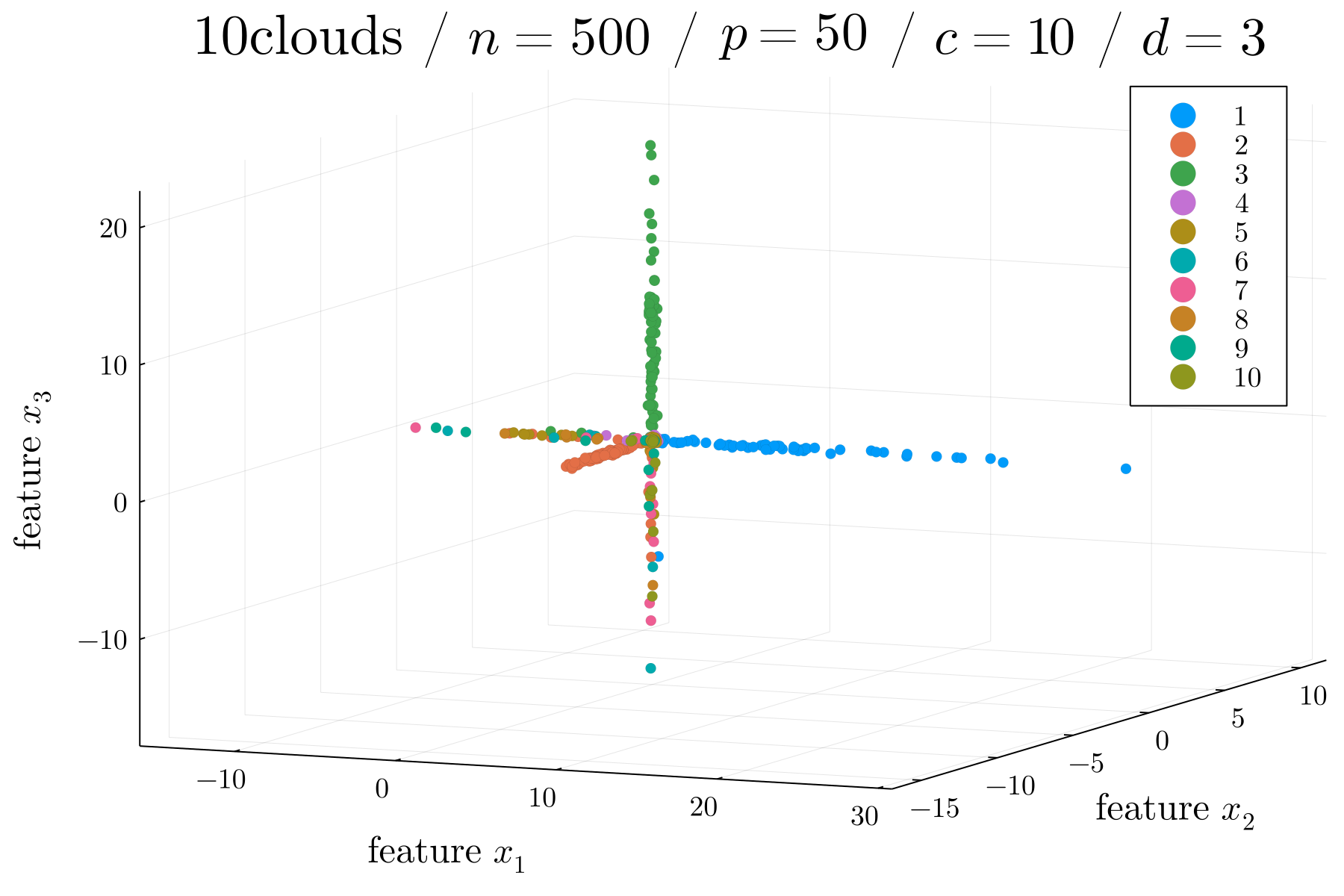}
    \caption{Representative subsets used in 5-fold cross validation for each simulated example. Here $n$, $p$, and $c$ correspond to numbers of samples, features, and classes, respectively.}
    \label{fig:simulation-examples}
\end{figure}

Figure~\ref{fig:simulation-examples} illustrates the shape of our simulated examples.

\section{Additional Details}
\label{ax:misc}

This section addresses additional details not covered in the preceding appendices.

\subsection[Class Vertices and Maximal Dead Zone Epsilon]{Class Vertices and Maximal Dead Zone $\epsilon$}

We prefer encoding $c$ class labels as vertices of a unit simplex with vertices
\begin{align*}
    \bv_{j} &= \begin{cases}
        (c-1)^{-1/2} \bone, & \text{if}~j=1 \\
        a\bone + b \be_{j-1}, & \text{if}~2 \le j \le c,
    \end{cases} \\
    a &= -\frac{1+\sqrt{c}}{(c-1)^{3/2}}
    \qquad
    b = \sqrt{\frac{c}{c-1}}.
\end{align*}
We use the maximal value of $\epsilon$ that avoids overlapping dead zones,
\[
    \epsilon = \frac{1}{2} \sqrt{\frac{2c}{c-1}}
\]

\subsection{Computing Environment}

All numerical examples are run on a Linux desktop (Manjaro, 5.10.89-1) equipped with 32 GB RAM and an Intel 10900KF processor locked at 4.9 GHz.
Only 8 cores are used.
Our algorithms are implemented in the Julia language, version 1.7.1.

\end{appendices}

\end{document}

%% file: preamble.tex
\def\sgn{\mathop{\rm sgn}\nolimits}

\def\tr{\mathop{\rm tr}\nolimits}

\def\dist{\mathop{\rm dist}\nolimits}

\def\prox{\mathop{\rm prox}\nolimits}
\def\argmin{\mathop{\rm argmin}\nolimits}

\newcommand{\bzero}{\boldsymbol{0}}
\newcommand{\bone}{\boldsymbol{1}}

\newcommand{\bb}{\boldsymbol{b}}

\newcommand{\be}{\boldsymbol{e}}

\newcommand{\br}{\boldsymbol{r}}

\newcommand{\bu}{\boldsymbol{u}}
\newcommand{\bv}{\boldsymbol{v}}

\newcommand{\bx}{\boldsymbol{x}}
\newcommand{\by}{\boldsymbol{y}}
\newcommand{\bz}{\boldsymbol{z}}
\newcommand{\bA}{\boldsymbol{A}}
\newcommand{\bB}{\boldsymbol{B}}
\newcommand{\bC}{\boldsymbol{C}}

\newcommand{\bE}{\boldsymbol{E}}
\newcommand{\bF}{\boldsymbol{F}}
\newcommand{\bG}{\boldsymbol{G}}

\newcommand{\bI}{\boldsymbol{I}}

\newcommand{\bN}{\boldsymbol{N}}

\newcommand{\bP}{\boldsymbol{P}}

\newcommand{\bU}{\boldsymbol{U}}
\newcommand{\bV}{\boldsymbol{V}}
\newcommand{\bW}{\boldsymbol{W}}
\newcommand{\bX}{\boldsymbol{X}}
\newcommand{\bY}{\boldsymbol{Y}}
\newcommand{\bZ}{\boldsymbol{Z}}

\newcommand{\bmu}{\boldsymbol{\mu}}

\newcommand{\bpsi}{\boldsymbol{\psi}}
\newcommand{\bsigma}{\boldsymbol{\sigma}}

\newcommand{\bSigma}{\boldsymbol{\Sigma}}

\newcommand{\Real}{\mathbb{R}}